\documentclass[11pt]{iopart}
\usepackage{bm}
\usepackage{hyperref}
\usepackage{amssymb}

\newcommand{\deriv}[2]{{\frac{\mathrm{d}{{#1}}}{\mathrm{d}{{#2}}}}}
\renewcommand{\d}{\mathrm{d}}
\newcommand{\pd}[2]{{\frac{\partial{{#1}}}{\partial{{#2}}}}}
\newcommand{\pnd}[3]{{\frac{\partial^{{#1}}{{#2}}}{\partial{{#3}}^{{#1}}}}}
\newcommand{\grad}{\nabla}
\newcommand{\R}{\mathcal{R}}
\newcommand{\vect}[1]{{\bm{\mathrm{{#1}}}}}

\renewcommand{\e}[1]{{\mathrm{e}^{{#1}}}}
\newcommand{\imag}{{\mathrm{i}}}

\newcommand{\X}{\mathcal{X}}
\newcommand{\Y}{\mathcal{Y}}

\newcommand{\Scl}{S_{\mathrm{cl}}}
\newcommand{\Scutoff}[1]{S_{{#1}}}
\newcommand{\op}{\mathcal{O}}

\newcommand{\dS}{\mathrm{dS}}

\renewcommand{\Re}{\mathrm{Re}\,}
\renewcommand{\Im}{\mathrm{Im}\,}
\newcommand{\symm}{\mathrm{sym}}
\newcommand{\genfunc}{Z}
\newcommand{\ren}{{\mathrm{ren}}}
\begin{document}
\title{Non-Gaussian Inflationary Perturbations from the dS/CFT Correspondence}
\date{\today}
\author{David Seery and James E. Lidsey}
\address{Astronomy Unit, School of Mathematical Sciences\\
  Queen Mary, University of London\\
  Mile End Road, London E1 4NS\\
  United Kingdom}
\eads{\mailto{D.Seery@qmul.ac.uk}, \mailto{J.E.Lidsey@qmul.ac.uk}}
\submitto{JCAP}
\begin{abstract}
We use the dS/CFT correspondence and bulk gravity to predict the form of the
renormalized holographic three-point correlation function of the operator
which is dual to the inflaton field perturbation during single-field,
slow-roll inflation.
Using Maldcaena's formulation of the correspondence, this correlator can be
related to the three-point function of the curvature perturbation generated
during single-field inflation, and we find exact agreement with
previous bulk QFT calculations. This provides a consistency check on
existing derivations of the non-Gaussianity from single-field inflation
and also yields insight into the nature of the dS/CFT correspondence.
As a result of our calculation, we obtain the properly renormalized dS/CFT
one-point function, including boundary contributions where derivative
interactions are present in the bulk.
In principle, our method may be employed to derive the $n$-point
correlators of the inflationary curvature perturbation within the context
of $(n-1)^{\mathrm{th}}$-order perturbation theory, 
rather than $n^{\mathrm{th}}$-order theory as in conventional approaches.

\vspace{3mm}
\begin{flushleft}
\end{flushleft}
\end{abstract}
\vfill
\maketitle

\section{Introduction}

\label{sec:1}

There has been considerable interest recently in understanding 
the nature of non-Gaussian features in the 
primordial curvature perturbation that is generated during 
early universe inflation
\cite{bartolo-matarrese-review,maldacena-nongaussian,seery-lidsey,
seery-lidsey-a,creminelli,lyth-rodriguez,lyth-rodriguez-a,acquaviva-bartolo,
lyth-zaballa,allen-gupta,zaballa-rodriguez,malik,
rigopoulos-shellard,rigopoulos-shellard-vantent,
calcagni-nongaussian,alabidi-lyth,
enqvist-jokinen,jokinen-mazumdar,barnaby-cline}.
This is motivated in part by the ever-increasing 
sensitivity of Cosmic Microwave Background (CMB)
data, through which one might hope to detect,
or at least set strong upper limits on, the primordial non-Gaussianity
\cite{wmap2006-params,wmap2006-temp,
mcewen-hobson,mukherjee-wang,vielva-martinez-gonzalez,
larsen-wandelt}. Moreover, from a theoretical perspective, 
the form of the primordial three-point correlation function may 
provide a more sensitive discriminant of inflationary microphysics 
than the tilt of the perturbation spectrum
\cite{alishahiha-silverstein,arkani-hamed-creminelli,seery-lidsey,
boubekeur-lyth,lyth-curvaton-ng,vernizzi-wands}.

Given these considerations, there is a pressing need to develop 
accurate theoretical techniques for calculating 
the primordial three-point function in concrete inflationary scenarios. 
An elegant method for determining the level of non-Gaussianity 
at horizon crossing in standard single-field, 
slow-roll inflation has been developed by 
Maldacena \cite{maldacena-nongaussian}. In this approach, 
the tree-level Feynman diagrams for an appropriate
vacuum-to-vacuum expectation value are evaluated. This technique  
was subsequently applied by various authors to other inflationary
scenarios, including models where higher-derivative operators are present in
the inflationary Lagrangian, or where more than one field is dynamically
important \cite{seery-lidsey,seery-lidsey-a,creminelli,
alishahiha-silverstein,arkani-hamed-creminelli}. 
It has been further demonstrated that the calculations can be extended 
beyond tree-level to include the effect of 
loop corrections \cite{weinberg-corrl}.

The purpose of the present paper is to develop an alternative 
method for calculating the three-point correlator for the inflaton field
perturbation that is based on the conjectured 
de Sitter/Conformal Field Theory (dS/CFT) 
correspondence \cite{strominger,strominger-inflation,maldacena-nongaussian}.
This correspondence states that quantum gravity
in four-dimensional de Sitter space is dual to a three-dimensional Euclidean 
CFT. In this picture, the timelike coordinate in de Sitter space is viewed 
as the scale parameter of the CFT and slow-roll inflation may be then
interpreted as a deformation of the CFT away from perfect scale invariance
\cite{strominger,klemm,leblond-myers}. 
It is natural, therefore, to investigate the relationship  
between the perturbations that are generated in the bulk inflationary 
physics and those of the holographically dual boundary field theory. 
The general rules for computing correlation functions in the dS/CFT framework
were presented by Maldacena \cite{maldacena-nongaussian} and shown to
produce the correct results in the case of a massless scalar field. 
The massive case was considered in \cite{larsen-schaar}, 
where the $\beta$-function and anomalous dimension in the dual CFT 
were related to the inflationary slow-roll parameters $\epsilon$ and $\eta$, 
which measure the logarithmic slope and curvature of the inflaton potential, 
respectively. The renormalized CFT generating functional 
that is dual to Einstein gravity coupled to a scalar field was
calculated by Larsen \& McNees \cite{larsen-mcnees}
(see also \cite{schaar}), who demonstrated that it correctly 
reproduces the amplitude and spectral tilt
of the density perturbation spectrum derived from standard bulk 
quantum field theory (QFT) calculations.  
(For reviews of such calculations, see, e.g.,
\cite{lidsey-liddle,liddle-lyth}).

The proposal that quantum gravity in de Sitter space is holographically
related to a CFT in one dimension fewer is motivated by the analogous
case of the AdS/CFT correspondence. A concrete realization of 
this correspondence is provided by perturbative type IIB string
theory on $\mathrm{AdS}_5 \times \bm{\mathrm{S}}^5$ with $N$
branes in the near-horizon limit. This can be related to an $\mathcal{N}=4$,
$SU(N)$ super-Yang Mills theory on the boundary of AdS
\cite{maldacena,witten-ads,aharony-gubser} (where $N \gg 1$;
for more details, see, e.g., \cite{johnson}). It is also possible to go
beyond this perturbative construction to more general scenarios.
Indeed, it has been suggested that the AdS/CFT correspondence 
may be viewed as a definition of what is meant by quantum gravity in a
space of asymptotically constant negative curvature \cite{witten-ds}.

However, the situation for the dS/CFT correspondence is less clear. 
There is presently no known concrete framework available 
which relates perturbative
string theory on dS space to some conformal field theory on the dS boundary.
Indeed, there may even be reasons to believe that such a correspondence
does not actually exist \cite{witten-ds,goheer-kleban,
dyson-kleban,dyson-lindesay}, at least for generic values of the Hubble
rate $H$ and Newton's constant $G$. Although the dS and AdS manifolds are
related by a double Wick rotation of the timelike and radial coordinates,
this property does not carry over to the
construction of holographic partners. We will encounter some of the
consequences of this in the analysis outlined below.

We will work with the more general version of the dS/CFT correspondence, which
states that \emph{any} gravitational theory in an asymptotically
dS space is holographically dual to some CFT
on the boundary. In this case, one can obtain information about the CFT by 
working entirely from the bulk theory and the assumed dS 
asymptotics. This is the approach taken in much of the AdS/CFT
literature \cite{maldacena-nongaussian,strominger,klebanov-witten,
muck-viswanathan,martelli-muck,bianchi-freedman-a,bianchi-freedman-b,
muck-prisco,bianchi-prisco-muck,haro-skenderis,kalkkinen-martelli,
fukuma-matsuura,
balasubramanian,skenderis} and in previous applications of the dS/CFT
correspondence to cosmology \cite{larsen-schaar,schaar,larsen-mcnees}.
Given that the precise form of the 
dS/CFT correspondence is at present unknown, we will  
compute the form that the holographic
correlation functions should take if the bulk 
calculation is to be recovered. If the holographic CFT that is dual to
bulk inflation is identified in the future, a direct 
calculation of its correlation functions will then be possible 
and this will allow a comparison to be made with the results 
of the present work. 
More specifically, we use Maldacena's formulation of the
dS/CFT correspondence \cite{maldacena-nongaussian}
to explicitly calculate the 
bulk prediction for the three-point correlator of the 
dual CFT which 
reproduces the primordial three-point correlator of the
inflaton field perturbation in single-field, slow-roll inflation. This 
provides a valuable consistency check of pre-existing results
and yields further insight into the nature of the dS/CFT correspondence 
itself and, more generally, into the properties of quantum gravity in de Sitter 
spacetime. 

The paper is organised as follows. We begin in Section \ref{sec:2}
by outlining those features of the dS/CFT correspondence 
that will be relevant for what follows and proceed in Section \ref{sec:3}
to present the bulk-to-boundary and bulk-to-bulk propagators for 
both free and interacting fields. This, together with the apparatus
of holographic renormalization \cite{nojiri-odintsov-holo,
martelli-muck,bianchi-freedman-b}, provides the necessary machinery
for calculating the holographic one-point function
$\langle \op \rangle$ of the operator $\op$ which is dual to the inflaton
$\varphi$. The derivation of $\langle \op \rangle$ 
is performed in Section \ref{sec:4}.
The argument is similar to that adopted in the AdS/CFT framework
\cite{muck-viswanathan,klebanov-witten,bianchi-freedman-b,martelli-muck},
but an extra subtlety arises due to the presence of derivative interactions
in the bulk. These interactions cause the appearance of non-zero boundary
terms at future infinity which can not necessarily be discarded, and this has 
important consequences when calculating the holographic one-point function.

Once the form of the one-point function has been determined, 
all other connected correlation functions can be derived directly from it
and we obtain the two- and three-point functions in Section \ref{sec:5}. 
Up to this point in the discussion, the analysis has been kept general, 
in the sense that the specific form of the interaction vertex has not been 
specified. In Section \ref{sec:6}, we discuss the effective 
field theory for the inflaton field during inflation, paying particular 
attention to the role of boundary terms in the action, 
and derive the corresponding holographic two- and three-point functions. 
We then proceed in Section \ref{sec:7} to demonstrate explicitly 
that the dS/CFT approach correctly leads to Maldacena's bulk QFT result 
for the primordial three-point non-Gaussianity in single-field inflation 
\cite{maldacena-nongaussian}. We also emphasize that  
the boundary terms that appear in the third-order 
bulk action for the inflaton perturbation can be accounted for by  
either performing a suitable field redefinition or by including 
such terms explicitly. Finally, we conclude with a discussion 
in Section \ref{sec:8}. 

\section{The dS/CFT Correspondence}

\label{sec:2}

In this section we briefly review the dS/CFT
\cite{strominger,dyson-lindesay,nojiri-odintsov-dscft,nojiri-odintsov-a}
conjecture,
in the variant proposed by Maldacena \cite{maldacena-nongaussian}.
The metric for the dS spacetime can be expressed in the form    
\begin{equation}
\label{dscft:metric}
  \d s^2 = -\d t^2 + a^2(t) \delta_{ij} \d x^i \d x^j =
           a^2 (\eta) [-\d\eta^2 + \delta_{ij} \d x^i \d x^j ] , 
\end{equation}
where $t$ denotes cosmic time, $\eta = \int \d t / a(t)$ defines 
conformal time and the scale factor $a = \e{Ht} = -(H \eta )^{-1}$.
In these coordinates, slices of constant $t$ are
manifestly invariant under the 3-dimensional Riemannian Poincar\'{e} group,
$ISO(3)$. The far past $(t \rightarrow -\infty)$ corresponds to 
$\eta \rightarrow  -\infty$ and the far future $(t \rightarrow \infty)$
to $\eta \rightarrow  0^-$. The boundary of dS space, 
$\partial \dS$, lies at $t =+ \infty$, together with
a point at $t = -\infty$, which makes the boundary compact \cite{witten-ads}.
The dual CFT can loosely be thought of as living on the Riemannian slice
at $\eta \rightarrow 0^-$, and inherits its symmetries.

The symmetries of dS space are broken in the presence of a 
scalar field, $\phi$, which propagates according to the field equation 
\begin{equation}
  \label{dscft:eqn}
  \ddot{\phi} + 3 H \dot{\phi} + \left( \frac{k^2}{a^2} + m^2 \right) \phi =
  - \deriv{V(\phi)}{\phi} ,
\end{equation}
where we have translated to Fourier space, $m$ denotes the mass 
of the field and possible higher-order interactions are parametrized 
by a potential $V(\phi)$. Even if $\phi$ evolves homogeneously,
the scale factor will be deformed owing
to the presence of energy--momentum in the bulk. Nonetheless, 
if the homogeneous part of the scalar field dominates as 
$t\rightarrow \infty$, the metric will be
asymptotically de Sitter such that $a(t) \sim \e{Ht}[1+
\Or(t^{-1})]$. In this case, provided $\phi$ is sufficiently light%
\footnote{If $m > 3H/4$, the conformal weight in Eq. \eref{dscft:delta}
becomes imaginary. This is one of the difficulties with the dS/CFT
proposal. As in Ref. \cite{maldacena-nongaussian}, we restrict our attention to
light fields, where this problem does not arise. In any event, only fields
satisfying $m < 3H/2$ are excited during a de Sitter epoch and could
generate a significant curvature perturbation in the late universe.},
it will behave near the boundary as
\begin{equation}
  \label{dscft:asymp}
  \phi \sim \hat{\phi} \e{\Delta_+ H t} (1 + \cdots)
  + \bar{\phi} \e{\Delta_- H t}(1 + \cdots) ,
\end{equation}
where
\begin{equation}
  \label{dscft:delta}
  \Delta_{\pm} = - \frac{3}{2} \pm \sqrt{\frac{9}{4} - \frac{m^2}{H^2}} .
\end{equation}
The boundary conditions determine the 
constants of integration $\{ \hat{\phi} , \bar{\phi}\}$. It follows that 
\begin{equation}
  \label{dscft:bc}
  \phi \sim \hat{\phi} \e{\Delta_+ H t}
\end{equation}
on the approach to $\partial\dS$ (as $t \rightarrow \infty$). Note 
that $\phi \rightarrow \hat{\phi}$ in the massless limit $m\rightarrow 0$.

Analogous to the AdS/CFT correspondence
\cite{maldacena,witten-ads,gubser-klebanov-polyakov,aharony-gubser,skenderis}, 
the dS/CFT correspondence is the formal statement that 
the wavefunction of quantum gravity on de Sitter space, $\Psi_{\dS}$, 
is given by the partition function of a dual CFT
\cite{strominger,maldacena-nongaussian}:
\begin{equation}
  \label{dscft:dscft-a}
  \genfunc_{CFT}[\hat{\phi}]  = \Psi_{\dS}
  \approx \e{\imag \Scl[\hat{\phi}]} , 
\end{equation}
where the second (approximate) equality holds when the curvature in the 
four-dimensional bulk spacetime is sufficiently small 
that the path integral arising in the definition 
of the wavefunction can be evaluated in the semi-classical limit. 
This yields the on-shell bulk action $\Scl[\hat{\phi}]$
evaluated on the classical solution. This action is a functional of the 
boundary data $\hat{\phi}$.

Eq.~\eref{dscft:dscft-a} implies that the
generating function of the CFT, $\genfunc_{CFT}[\hat{\phi}]$, must also
depend on $\hat{\phi}$. However, a generating function is typically
only a functional of the sources in the theory, since all dynamical fields
are integrated out. Therefore, it is natural to interpret $\hat{\phi}$
as the source for some operator $\op$ which is dual,
up to a constant of proportionality, to $\phi$ under the dS/CFT correspondence. 
This implies that Eq. \eref{dscft:dscft-a} can be expressed as
\cite{maldacena,witten-ads,muck-viswanathan,skenderis,
balasubramanian-boer-a,balasubramanian-boer-b}
\begin{equation}
  \label{dscft:dscft}
  \left\langle \exp \left( \int_{\partial \dS}
  \d^3 x \; \hat{\phi} \op \right)
  \right\rangle_{CFT} =
  \genfunc_{CFT}[\hat{\phi}] \approx
  \exp\left( \imag \Scl[\hat{\phi}] \right) ,
\end{equation}
where $\d^3 x$ is an invariant volume measure on the boundary $\partial \dS$.
Eq. \eref{dscft:dscft} is the statement of the dS/CFT correspondence 
that we employ in this paper, although in practice 
it must be supplemented with counterterms, as discussed 
in Section~\ref{sec:4}. 
The CFT correlators in the absence of the source $\hat{\phi}$
can be recovered by functionally differentiating 
Eq. \eref{dscft:dscft} with respect to the source, $\hat{\phi}$, and setting 
$\hat{\phi}=0$ after the differentiation. Thus,  
\begin{equation}
  \label{dscft:taylor}
  \langle \op(\vect{x}_1) \cdots \op(\vect{x}_n) \rangle =
  \left. \frac{\delta^n \ln \genfunc_{CFT}[\hat{\phi}]}
  {\delta \hat{\phi}(\vect{x}_1)
  \cdots \delta \hat{\phi}(\vect{x}_n)} \right|_{\hat{\phi} = 0} .
\end{equation}

A physical interpretation may also be given to the other 
constant $\bar{\phi}$ that arises in the asymptotic solution 
\eref{dscft:asymp}. In the AdS/CFT correspondence, 
this parameter is identified, modulo a constant of proportionality,
as the vacuum expectation value (VEV) $\langle \op \rangle$ 
\cite{witten-ads,klebanov-witten,haro-skenderis}. 
In general, the numerical value of the constant of proportionality 
depends on the specific theory under investigation and must be evaluated 
by direct calculation. Since $\hat{\phi}$ is interpreted as a source, 
the quantity $\bar{\phi}$ is often referred to as the 
\emph{response} in the AdS/CFT literature. 
We will show that in the dS/CFT correspondence, the identification 
$\langle \op \rangle \propto \bar{\phi}$ must be
modified by including boundary terms that arise from interactions 
in the bulk field.  We will also explicitly determine 
the constant of proportionality in Section~\ref{sec:4.1}.

The interpretation of $\hat{\phi}$ as a source for $\op$ and $\bar{\phi}$
as its VEV has an analogue in the bulk theory 
\cite{minces-rivelles,balasubramanian}. 
This serves to make the identifications more transparent.
The $\Delta_-$ solution in Eq. \eref{dscft:asymp}
decays near the boundary and is normalizable. 
It corresponds to a finite energy excitation of the bulk
theory, just as the acquisition of a VEV by $\op$ is a finite energy
excitation of the boundary CFT. The $\Delta_+$ solution, 
on the other hand, is not normalizable
and corresponds to an infinite energy excitation of the bulk theory. It 
should therefore be viewed as a deformation of the gravitational background.
This is equivalent to the deformation of the CFT Lagrangian by the operator
$\hat{\phi} \op$. 

\section{Bulk-to-Boundary and Bulk-to-Bulk Propagators}

\label{sec:3}

The above discussion implies that a key feature of employing 
the dS/CFT correspondence is the Dirichlet problem in de Sitter 
space, i.e., the problem of finding the solution to the bulk field equation 
\eref{dscft:eqn} subject to the boundary condition 
that $\phi \rightarrow \hat{\phi}$ (for the massless case) 
in the far future. In this section, we identify the solution
that satisfies this property.
This will enable us to calculate the response $\bar{\phi}$ in 
terms of the source $\hat{\phi}$.

If the metric is asymptotically dS as $t \rightarrow \infty$,  
the self-interaction described by the potential $V(\phi)$ 
in Eq. (\ref{dscft:eqn}) becomes unimportant. 
In this limit, the general solution to Eq. (\ref{dscft:eqn}) is 
given in terms of Bessel functions: 
\begin{equation}
\label{gensol}
  \phi =   (-k\eta )^{3/2}J_{\nu} (-k \eta) \bar{\phi} +
  (-k\eta)^{3/2} Y_{\nu} (-k\eta) \hat{\phi}  ,
\end{equation}
where \cite{breitenlohner-freedman-a,breitenlohner-freedman-b,
mezincescu-townsend}
\begin{equation}
  \nu^2 = \frac{9}{4} - \frac{m^2}{H^2} =
  \left( \Delta_{\pm} + \frac{3}{2} \right)^2  .
\end{equation}
One of the boundary conditions that can be imposed is 
that the field should be in its vacuum state in the asymptotic past 
when it is deep inside the de Sitter horizon. The usual choice is to 
invoke the Bunch--Davies vacuum \cite{birrell-davies}, 
which corresponds to specifying the solution in terms of 
a Hankel function of the second kind or order $\nu$:
\begin{equation}
\label{hankelprop}
\phi \propto (-k\eta)^{3/2} H_\nu^{(2)}(-k\eta)  .
\end{equation}

In the following Subsections, we determine the constant of proportionality 
in Eq. (\ref{hankelprop}) for free and interacting fields, 
respectively.   

\subsection{Free Fields}

\label{sec:3.1}

A free field has $m=V=0$. As discussed above, the second boundary condition 
that should be imposed is the requirement 
that $\phi \rightarrow \hat{\phi}$ at
the boundary $\partial {\rm dS}$. This boundary condition 
may be satisfied by identifying a boundary Green's function
$K(\eta;k)$ that is a solution to the field equation \eref{dscft:eqn} 
and obeys the condition $K \rightarrow 1$ on 
$\partial\dS$ \cite{morse-feshbach}.
In coordinate space, this is equivalent to the requirement 
that $K$ approach a $\delta$-function on the boundary
and the appropriate solution is given by
\begin{equation}
\label{twopt:green}
  K(\eta,k) 
  = \imag \left( \frac{k}{2} \right)^{3/2} \Gamma(-1/2) (-\eta)^{3/2} 
  H_{3/2}^{(2)} (-k\eta) = (1-ik\eta) e^{ik \eta} .
\end{equation}
The function $K$ is sometimes referred to as the `bulk-to-boundary 
propagator' and its momentum dependence is entirely 
specified by $k \equiv |\vect{k}|$. 

The solution for the field $\phi$ which obeys
the boundary condition \eref{dscft:bc} may now be written down 
immediately:  
\begin{equation}
  \label{twopt:soln}
  \phi(\eta,\vect{k}) = K(\eta,k) \hat{\phi}(\vect{k}) .
\end{equation}
The response $\bar{\phi}$ may be identified in terms of the source
$\hat{\phi}$ by expanding solution \eref{twopt:soln} 
as an asymptotic series near $\eta \approx 0^-$ and comparing the 
coefficients of the expansion with the corresponding expansion of the general 
solution \eref{gensol}. More specifically, the asymptotic form of the 
solution \eref{gensol} for a massless scalar field behaves 
quite generally near future infinity as 
\begin{eqnarray}
  \label{dirichlet:asympt}
  \phi \sim \hat{\phi} + \hat{\phi} \lambda \eta^2 +
  \bar{\phi} \eta^3 + \cdots ,
\end{eqnarray}
where `$\cdots$' denotes terms of $\Or(\eta^4)$ or higher and $\lambda$ 
is a constant, coming from the subleading term of the $\Delta_+$ mode.
It follows from Eq. \eref{dirichlet:asympt} 
that the coefficient of the $\eta^3$ term 
in the near-boundary expansion of $\phi$ is indeed the response $\bar{\phi}$.
Hence, expanding the solution \eref{twopt:soln} near the boundary,
where $\eta \approx 0^-$, and evaluating the coefficient 
of the $\eta^3$ term, implies that the response is given by  
\begin{equation}
  \label{twopt:response}
  \bar{\phi}(\vect{k})
  = \imag \frac{k^3}{3} \hat{\phi}(\vect{k}) .
\end{equation}

\subsection{Interacting Fields}

\label{sec:3.2}

This formalism can be extended to interacting fields. Within the context
of the AdS/CFT correspondence, interactions were introduced in
Ref. \cite{muck-viswanathan}
and the first connected correlation functions containing three or
more fields were derived in \cite{bianchi-marchetti,bianchi-prisco-muck,
muck-prisco}.  

An interacting field implies that the potential $V$ in Eq. \eref{dscft:eqn}
must contain cubic or higher-order couplings with some coupling constant,
$g$. Our motivation for considering interacting fields 
arises from the possibility that primordial non-Gaussianities,
which would be sourced by inflaton interactions,
may soon be detectable in the CMB temperature anisotropy power spectrum. 
We will restrict our discussion to cubic 
interactions, since these are of most relevance to inflationary cosmology.  
To be specific, we consider a bulk action of the form  
\begin{equation}
  \label{onept:action}
  S = \int \d \eta \, \d^3 x \; \left[ \frac{1}{2} a^2 \left(
  \phi'^2
  - ( \partial \phi )^2 \right) + L_3 \right] ,
\end{equation}
where $L_3$ denotes interaction terms that are 
of order $\phi^3$ with coupling $g$. (These are given later by Eq. 
\eref{infl:phi} for the inflationary scenario). 
Without loss of generality, we assume that no boundary terms are
present in Eq. \eref{onept:action}. If this is not the case, such terms
can be removed, either by a field redefinition or by including a 
total derivative in the bulk interaction term $L_3$. 

The corresponding field equation for $\phi$ follows after varying 
Eq. \eref{onept:action}: 
\begin{equation}
  \label{onept:feq}
  \phi'' + 2 \frac{a'}{a} \phi' - \partial^2 \phi = \frac{1}{a^2}
  \frac{\delta L_3}{\delta \phi} ,
\end{equation}
where a prime denotes differentiation with respect to conformal time
and $\delta L_3/\delta \phi$ is defined by the rule
\begin{equation}
  \label{onept:deltaLrule}
  \int \d \eta \, \d^3 x \; \delta L_3 = \int \d \eta \, \d^3 x \;
  \frac{\delta L_3}{\delta \phi}\delta\phi + \int_\partial
  \d^3 x \; \xi_2 \delta\phi ,
\end{equation}
in which the boundary term $\xi_2$ is generated by the 
integrations by parts that may be needed in order 
to cast $\delta L_3$ in the form $(\delta L_3/\delta\phi)
\delta\phi$. In calculating the field equation, we are free to choose the
boundary conditions for $\delta\phi$ so that it vanishes on the boundary.
As a result, $\xi_2$ does not contribute to Eq. \eref{onept:feq}. However,
as we shall see, this does \emph{not} necessarily imply
that such a term is irrelevant: although it has no consequences for the
field equations, $\xi_2$ plays a vital role in determining the
correct 1-point function, $\langle \op \rangle$.
Our treatment improves the analysis of M\"{u}ck \& Viswanathan
\cite{muck-viswanathan}, who allowed for the possibility of
arbitrary interactions in the AdS/CFT framework but took the boundary term
described by $\xi_2$ to generate a contribution to the $\phi$ field equation
(or implicitly allowed a modification to the action which would cancel
any such terms). We discuss this point in more detail below.

The presence of interaction terms implies that 
the field equation, Eq~\eref{onept:feq}, can not be solved exactly. 
As a result, we make the standard assumption that $|g| \ll 1$ and 
then perfom a calculation in perturbation theory around $g$.
Within the context of the inflationary scenario, $g$ will be of order
$\dot{\phi}/H$, and this assumption is equivalent to  invoking the
slow-roll approximation.
We denote by a subscript `$n$' terms that are of order $g^n$. 
Thus, at $\Or(g^0)$, Eq. \eref{onept:feq} reduces to the free field case
\begin{equation}
  \label{threept:zero}
  \phi_0'' - \frac{2}{\eta} \phi'_0 + k^2 \phi_0 = 0 
\end{equation}
and the solution to Eq. \eref{threept:zero} that satisfies the
Dirichlet boundary condition $\phi_0 \rightarrow \hat{\phi}$ 
on $\partial \dS$ is given by the boundary Green's function
\eref{twopt:soln} with $\phi$ replaced by $\phi_0$.

At the next order, 
Eq. \eref{onept:deltaLrule} can be solved using the bulk Green's function
$G(\eta,\tau;k)$, which depends only on $k = |\vect{k}|$, and satisfies
\begin{equation}
  \left( \pnd{2}{}{\eta} - \frac{2}{\eta} \pd{}{\eta} + k^2 \right)
  G(\eta,\tau;k) = \delta(\eta-\tau) .
\end{equation}
The boundary conditions for $G$ should be consistent with the
dS/CFT interpretation. Specifically, $G$ should be regular in the deep
interior of spacetime, where $\eta \rightarrow -\infty$. Near the boundary,
$\eta \uparrow 0$, we require  $G \rightarrow 0$, so that 
the $\Or(g^1)$ corrections to $\phi_0$
do not violate the boundary condition $\phi \rightarrow \hat{\phi}$. 
The solution satisfying these conditions is 
\begin{equation}
  \label{threept:bulk-green}
  G(\eta,\tau;k) = \frac{\pi \imag}{2} (-\eta)^{3/2} (-\tau)^{-1/2}
  \times \left\{
  \begin{array}{ll}
    H_{3/2}^{(2)}(-k\tau) J_{3/2}(-k\eta), & \eta > \tau \\
    H_{3/2}^{(2)}(-k\eta) J_{3/2}(-k\tau), & \eta < \tau .
  \end{array}
  \right.
\end{equation}

Hence, combining the $\Or(g^1)$ solution \eref{threept:bulk-green}
with the $\Or(g^0)$ solution \eref{twopt:green} yields the result
\begin{equation}
  \label{onept:cs}
  \fl
  \phi(\eta,\vect{x})
  = \int \frac{\d^3 k}{(2\pi)^3} \; K(\eta;k) \hat{\phi}(\vect{k})
  \e{-\imag \vect{k}\cdot\vect{x}} +
  \int \frac{\d^3 k}{(2\pi)^3} \, \d\tau \; G(\eta,\tau;k)
  \frac{1}{a^2} \widetilde{\frac{\delta L_3}{\delta\phi(\tau,\vect{k})}}
  \e{-\imag \vect{k}\cdot\vect{x}} ,
\end{equation}
where a tilde denotes the Fourier transform
\begin{equation}
  \widetilde{\frac{\delta L_3}{\delta \phi(\tau,\vect{k})}} =
  \int \d^3 x \; \e{\imag \vect{k}\cdot\vect{x}} \;
  \frac{\delta L_3}{\delta \phi(\tau,\vect{x})} ,
\end{equation}
and the $\tau$ integral in \eref{onept:cs} is performed along a contour which
is slightly rotated for large $|\eta|$,
i.e., $\eta \mapsto \eta(1-\imag\delta)$. In the remainder of this paper,
integrals over conformal time such as Eq.~\eref{onept:cs}
will be written merely as $\int_{-\infty}^0
\d \tau$ with this convention understood. This rotation,
which is already implicit in Eq.~\eref{onept:action}, is equivalent to the
projection onto the interacting quantum vacuum \cite{maldacena-nongaussian}
which appears in bulk QFT calculations of the three-point function
\cite{seery-lidsey,peskin-schroeder}. One might be tempted to imagine
that this procedure, which is tantamount to dropping a rapidly oscillating
term for $\eta \rightarrow -\infty$, corresponds to a renormalization
prescription. However, this is not the case. Since it occurs in the field
theory infra-red (equivalently, the gravitational ultra-violet)
it has nothing to do with renormalization.
The dangerous divergences which must be regulated all appear in the
field theory ultra-violet (equivalently, the gravitational infra-red),
where $\eta \approx 0^-$, and are safely removed by the holographic
renormalization procedure to be discussed in Section~\ref{sec:4}.

It will be important in what follows that in the vicinity of the 
boundary, where $\eta \simeq 0^-$, the Green's function 
\eref{threept:bulk-green} behaves asymptotically as 
\begin{equation}
  \label{threept:green-boundary}
  G(\eta,\tau;k) \simeq \frac{\eta^3}{3\tau^2} K(\tau;k) + \cdots ,
\end{equation}
where the dots denote higher-order terms in conformal time.
This near-boundary behaviour of the Green's function can
be employed to read off the coefficient of $\eta^3$ in the solution
\eref{onept:cs} for small $\eta$. 
As discussed above, this coefficient is the response $\bar{\phi}$.
The contribution to the response from the bulk-to-boundary propagator
in \eref{onept:cs} is given by Eq. \eref{twopt:response}, 
whereas the contribution from the bulk-to-bulk propagator 
in the limit of small $\eta$ is
\begin{equation}
  \label{threept:phibar-bdy}
  \Delta\bar{\phi}(\eta,\vect{k})
  \simeq
  \frac{\eta^3}{3}
  \int_{-\infty}^{\eta} \frac{\d \tau}{a^2 \tau^2} \;
  K(\tau;k) \widetilde{\frac{\delta L_3}{\delta \phi(\tau,\vect{k})}}
  + \cdots ,
\end{equation}
where we have written the integral explicity only over the $\tau < \eta$
branch of \eref{threept:bulk-green}. The other branch, denoted `$\cdots$',
where $\tau > \eta$ and other higher-order terms in $\eta$ contribute only
at $\Or(\eta^4)$, and can be safely ignored for the purposes
of calculating $\bar{\phi}$.

It follows, therefore, that the response for the interacting field is given 
by 
\begin{equation}
  \label{onept:phibar}
  \bar{\phi}(\vect{k}) = \imag \frac{k^3}{3} \hat{\phi}(\vect{k}) +
  \frac{H^2}{3} \int^0_{-\infty} 
  \d \tau \; K(\tau; k) \widetilde{\frac{\delta L_3}
  {\delta \phi(\tau,\vect{k})}} ,
\end{equation}
where we have set the upper limit of integration to be
zero rather than $\eta$, since the difference is of higher
order in $\eta$ and therefore irrelevant when calculating the
response.

Note that the calculation in this section has been purely formal,
without regard to the convergence of the $\tau$ integral in
Eq.~\eref{onept:phibar}. In order to assign a precise meaning to this
purely formal expression, one must carry out a renormalization procedure
to remove possible infinities near the boundary surface $\eta \approx 0^-$.
In the following Section, we proceed to calculate the one-point function of
the deformed CFT in terms of the response \eref{onept:phibar}, taking
holographic renormalization into account.

\section{The Holographic One-point Function}

\label{sec:4}

Eq. \eref{dscft:dscft} cannot be precisely correct as it stands, 
since both sides are \emph{a priori} divergent. On the gravitational side, 
the on-shell action $\Scl[\hat{\phi}]$ exhibits an infra-red 
divergence as $t \rightarrow \infty$. On the CFT side, one must expect
the usual ultra-violet divergences of any local quantum field theory to be
present.
These divergences can be removed by the addition of
appropriate counterterms, which should be understood to be included in
Eq. \eref{dscft:dscft}. 

In order to determine the nature 
of these counterterms, one regularizes the
gravitational action by introducing a cut-off in the spacetime at some
large, but finite, value of $t$, corresponding to a very small negative
value of conformal time, $\eta = -\epsilon$, where $0 < \epsilon \ll 1$. 
We will denote the on-shell action computed in this regularized 
spacetime by $\Scutoff{\epsilon}$. This action diverges as 
$\epsilon \rightarrow 0$, but after the divergences have been cancelled 
by the counterterms and the regularization removed, a finite 
contribution will remain. This remainder is interpreted as the renormalized CFT 
generating functional.

One might worry that this cut-off procedure is coordinate dependent, and that
the results might change if the cut-off was calculated using a different
choice of coordinate time. Our ability to reparametrize the spacetime
cut-off corresponds to the insensitivity of the CFT to the regulator which
is chosen. Moreover, although this subtraction scheme appears to explicitly
break Lorentz covariance, it turns out that the counterterms can always
be rewritten covariantly \cite{bianchi-freedman-a,martelli-muck}.
Hence, if we calculate physical quantities with 
the specific regulator $\eta = -\epsilon$, the scheme-invariant
quantities (such as observables of physical interest) should coincide with
those calculated using any other regularization scheme.

\subsection{The Holographic One-point Function in terms of the Response}

\label{sec:4.1}

It remains to explicitly calculate the one-point 
function $\langle \op(\vect{k}) \rangle_{\hat{\phi}}$. This will 
demonstrate the holographic renormalization procedure in action and
allow us to determine the constant of proportionality in the relation
$\langle \op \rangle \propto \bar{\phi}$.
The argument is similar to Ref. \cite{muck-viswanathan}, but care must be
taken to include the effect of boundary terms.

We first integrate by parts in the quadratic sector of
Eq. \eref{onept:action}, after which one obtains
\begin{eqnarray}
  \fl
  \Scutoff{\epsilon} = \int_{\eta=-\epsilon} \d^3 x \; \left[ \frac{1}{2}
  a^2 \phi \phi' \right] -
  \int_{\eta=-\infty}^{\eta=-\epsilon}
  \d \eta \, \d^3 x \; \left[ \frac{1}{2} \phi (a^2\phi')' -
  \frac{1}{2} a^2 \phi \partial^2 \phi \right] \nonumber \\
  +\int_{\eta=-\infty}^{\eta=-\epsilon} \d\eta \, \d^3 x \; L_3 ,
\end{eqnarray}
where the first term is evaluated on the slice corresponding to
$\eta = -\epsilon$. It then follows from 
Eq. \eref{onept:feq} that the on-shell action takes the form 
$\Scutoff{\epsilon} = \Scutoff{\epsilon|1} + \Scutoff{\epsilon|2} +
\Scutoff{\epsilon|3}$, where
\begin{eqnarray}
  \nonumber
  \Scutoff{\epsilon|1} = \int_{\eta=-\epsilon}
  \d^3 x \; \left[ \frac{1}{2} a^2 \phi \phi' \right] \\
  \nonumber
  \Scutoff{\epsilon|2} = -
  \int_{\eta=-\infty}^{\eta=-\epsilon} \d\eta \,
  \d^3 x \; \frac{1}{2} \phi \frac{\delta L_3}{\delta \phi} \\
  \label{onept:interim}
  \Scutoff{\epsilon|3} =
  \int_{\eta=-\infty}^{\eta=-\epsilon} \d\eta \, \d^3 x \; L_3 .
\end{eqnarray}  

Eqs. \eref{dscft:dscft} and \eref{dscft:taylor} imply 
that the one-point function is determined by the variation 
$\delta \Scutoff{\epsilon}/\delta \hat{\phi}$ in the limit $\epsilon
\rightarrow 0$, and this may be evaluated 
by considering the variation of each of the terms in  
Eq. \eref{onept:interim} separately. 
The first term is a surface integral evaluated 
at small $\eta$ and the expansion given in Eq. \eref{dirichlet:asympt}
implies that this can be expressed in the form
\begin{eqnarray}
  \label{dscft:cut-offs}
  \Scutoff{\epsilon|1} \simeq
  \int_{\eta=-\epsilon} \d^3 x \;
  \left[ - \frac{1}{\epsilon} \frac{\hat{\phi}^2 \lambda}{H^2} +
  \frac{3}{2} \frac{\hat{\phi} \bar{\phi}}{H^2} + \Or(\epsilon) \right] .
\end{eqnarray}
The first term in Eq. \eref{dscft:cut-offs} diverges 
as $\epsilon \rightarrow 0$ and should therefore be
subtracted by an appropriate counterterm. The second term is finite,
whereas the remaining terms are all $\Or(\epsilon)$ and 
therefore vanish as the regulator is removed. After substitution of 
Eq. \eref{onept:phibar}, therefore, this contribution to the 
on-shell action reduces to 
\begin{eqnarray}
  \fl
  \Scutoff{\epsilon|1}^\ren \simeq
  \int \frac{\d^3 k_1 \, \d^3 k_2}{(2\pi)^3} \;
  \delta(\vect{k}_1 + \vect{k}_2) \Bigg[ \frac{\imag k_1^3}{2H^2}
  \hat{\phi}(\vect{k}_1) \hat{\phi}(\vect{k}_2)
  \nonumber \\ 
  \mbox{} + \frac{1}{2}
  \int_{-\infty}^0 \d \tau \; K(\tau;k_2) \hat{\phi}(\vect{k}_1)
  \widetilde{\frac{\delta L_3}{\delta\phi(\tau,\vect{k}_2)}} \Bigg] ,
\end{eqnarray}
where we write $\simeq$ to denote expressions that are valid 
up to terms which vanish as $\epsilon \rightarrow 0$.
Varying with respect to $\hat{\phi}(\vect{k})$ then 
implies that\footnote{A potentially subtle point is that 
no surface terms are ever generated from
a variation with respect to the \emph{boundary} field, i.e., variations
of the form $\delta/\delta \hat{\phi}$, even when applied to
derivatives of $\phi$. For example,
\begin{equation}
  \label{onept:surf}
  \frac{\delta}{\delta \hat{\phi}(\vect{k})}
  \frac{\partial^n}{\partial \eta^n} \phi(\eta,\vect{p}) =
  \frac{\delta}{\delta \hat{\phi}(\vect{k})}
  \frac{\partial^n}{\partial \eta^n}K(\eta;\vect{p}) \hat{\phi}(\vect{p}) =
  \frac{\partial^n}{\partial \eta^n}K(\eta;\vect{p}) \delta(\vect{p}-\vect{k}) .
\end{equation}
Surface terms only ever arise from variations with respect to the \emph{bulk}
field $\phi(\eta,\vect{x})$, rather than $\hat{\phi}(\vect{x})$.}
\begin{eqnarray}
  \nonumber\fl
  \label{firstvary}
  \frac{\delta \Scutoff{\epsilon|1}^\ren}{\delta \hat{\phi}(\vect{k})} \simeq
  \frac{1}{(2\pi)^3} \frac{\imag k^3}{H^2} \hat{\phi}(-\vect{k}) +
  \frac{1}{2} \frac{1}{(2\pi)^3} \int_{-\infty}^0 \d \tau \; K(\tau;k)
  \widetilde{\frac{\delta L_3}{\delta \phi(\tau,-\vect{k})}} \\
  \mbox{} + \frac{1}{2} \int_{-\infty}^{0}
  \d \tau \, \frac{\d^3 p}{(2\pi)^3} \;
  K(\tau;p)
  \hat{\phi}(\vect{p}) \frac{\delta}{\delta \hat{\phi}(\vect{k})}
  \widetilde{\frac{\delta L_3}{\delta\phi(\tau,-\vect{p})}} .
\end{eqnarray}

We also require the variations of the other terms in 
Eq. \eref{onept:interim}. After translating to Fourier space, 
the variation of $\Scutoff{\epsilon|2}$ is given by 
\begin{eqnarray}
  \nonumber\fl
  \label{secondvary}
  \frac{\delta \Scutoff{\epsilon|2}}{\delta \hat{\phi}(\vect{k})} =
  - \frac{1}{2}\frac{1}{(2\pi)^3} \int_{-\infty}^{-\epsilon}
  \d \tau \; K(\tau;k)
  \widetilde{\frac{\delta L_3}{\delta\phi(\tau,-\vect{k})}} \\
  \mbox{} - \frac{1}{2} \int_{-\infty}^{-\epsilon}
  \d \tau \, \frac{\d^3 p}{(2\pi)^3}
  \; K(\tau,p) \hat{\phi}(\vect{p}) \frac{\delta}{\delta
  \hat{\phi}(\vect{k})} \widetilde{\frac{\delta L_3}{\delta\phi(
  \tau,-\vect{p})}} .
\end{eqnarray}
On the other hand, one can show by employing 
rule \eref{onept:deltaLrule} (which expresses how
$\delta L_3$ is related to $\delta L_3/\delta \phi$ and the surface term
$\xi_2$) that $\delta \Scutoff{\epsilon|3}$ behaves like
\begin{equation}
\label{thirdvary}
  \frac{\delta \Scutoff{\epsilon|3}}{\delta \hat{\phi}(\vect{k})} =
  \frac{1}{(2\pi)^3} \xi_2(-\epsilon,-\vect{k}) K(-\epsilon,k) +
  \int_{-\infty}^{-\epsilon} \frac{\d \tau}{(2\pi)^3} \;
  K(\tau;k) \widetilde{\frac{\delta L_3} {\delta\phi(\tau,-\vect{k})}} .
\end{equation}
Finally, after collecting together 
Eqs. \eref{firstvary}, \eref{secondvary} and 
\eref{thirdvary}, we find that 
\begin{equation}
  \fl\label{totalvary}
  \frac{\delta\Scutoff{\epsilon}}{\delta\hat{\phi}(\vect{k})} \simeq
  \frac{1}{(2\pi)^3}\frac{\imag k^3}{H^2} \hat{\phi}(-\vect{k}) +
  \int_{-\infty}^{-\epsilon} \d\tau \; K(\tau,k)
  \widetilde{\frac{\delta L_3}{\delta\phi(\tau,-\vect{k})}} +
  \frac{1}{(2\pi)^3} \xi_2(-\epsilon,-\vect{k})K(-\epsilon,k) .
\end{equation}

In order to take the $\epsilon \rightarrow 0$ limit, we
define a renormalized response, $\bar{\phi}^{\ren}$, such that 
\begin{equation}
  \fl
  \label{renresponse}
  \bar{\phi}^\ren(\vect{k}) = \lim_{\epsilon \rightarrow 0} \left(
  \imag \frac{k^3}{3} \hat{\phi}(\vect{k})
  + \frac{H^2}{3} \int_{-\infty}^{-\epsilon}
  \d\tau \; K(\tau,k) \widetilde{\frac{\delta L_3}{\delta(\tau,\vect{k})}}
  + \mbox{counterterms} \right) ,
\end{equation}
where the counterterms are chosen so that the limit exists
\cite{bianchi-freedman-b}. The usual 
ambiguities arise when considering finite counterterms,  
which are independent of $\epsilon$. These counterterms
shift the final value of $\langle \op(\vect{k}) \rangle$,
but since they are renormalization-scheme dependent, 
we can always choose a scheme in which they are absent.
This is the ``minimal'' subtraction prescription of Ref.
\cite{bianchi-freedman-b}.
Applying the same procedure to $\xi_2$ yields a renormalized surface term:
\begin{equation}
  \label{renboundary}
  \xi_2^\ren(\vect{k}) = \lim_{\epsilon \rightarrow 0}
  \left( \xi_2(-\epsilon,\vect{k}) K(-\epsilon,k) + \mbox{counterterms}
  \right) .
\end{equation}
Hence, the correctly renormalized
one-point function can be written in the simple and explicit form
\begin{equation}
\label{dscft:onept}
  \langle \op(\vect{k}) \rangle_{\hat{\phi}} = \imag
  \frac{\delta \Scutoff{\epsilon}}{\delta \hat{\phi}(\vect{k})} =
  \frac{1}{(2\pi)^3} \frac{3 \imag}{H^2}
  \bar{\phi}^\ren(-\vect{k}) + \frac{\imag}{(2\pi)^3} \xi_2^{\ren}(-\vect{k}) .
\end{equation}
The presence of counterterms ensures that the integral
in Eq.~\eref{renresponse} is finite. This should 
be compared with the bare response, Eq.~\eref{onept:phibar},
which may contain divergences at future infinity. (In general, 
there are no divergences arising from the asymptotic past 
\cite{weinberg-corrl}). 

Eqs.~\eref{renresponse}--\eref{renboundary} can be used to obtain the
holographic counterterms explicitly. However, such terms are always 
found to be imaginary in models that have been studied to date 
and Maldacena has conjectured that this property should 
hold in general \cite{maldacena-nongaussian}.
In this case, it follows that since
the interesting component of the one-point function is purely real, 
the correctly renormalized one-point function may be derived 
by dropping any terms in Eq.~\eref{totalvary} which give rise to an
imaginary part in $\langle \op(\vect{k}) \rangle$. This prescription is
simpler to use in practice, although it should be emphasized 
that Eqs.~\eref{renresponse}--\eref{renboundary} 
can always be employed to compute the counterterms
without making any prior assumptions about their complex nature.

Eq.~\eref{dscft:onept} differs from results previously obtained 
in the AdS/CFT context \cite{muck-viswanathan,bianchi-freedman-b,
martelli-muck,bianchi-freedman-a}
due to the presence of the surface contribution $\xi_2$.
This term arises because, as is usual in field theory in de Sitter space,
we have discarded boundary terms at future infinity when calculating the
field equations.
This is the usual situation. In Ref. \cite{muck-viswanathan},
M\"{u}ck \& Viswanathan assumed that the
interaction term was written as $I_{\mathrm{int}}$, which is equal to
$\int \d^4 x \; L_3$ in our notation, and the field equations were taken to be
$-\grad^a \grad_a \phi = \delta I_{\mathrm{int}}/\delta \phi$. As a result,
M\"{u}ck \& Viswanathan derived Eq.~\eref{dscft:onept} with $\xi_2 = 0$.
There is no discrepancy, because if $L_3$ does not contain derivative terms
then $\xi_2$ is always zero, whereas if such terms are present then
$\delta I_{\mathrm{int}}/\delta \phi$ implicitly contains a $\delta$-function
term at the boundary which would replicate the effect of $\xi_2$.
As a result, our analysis is consistent with previous AdS/CFT
computations which did not include derivative interactions
\cite{martelli-muck,bianchi-freedman-a,bianchi-freedman-b,skenderis}.

To summarize thus far, we have calculated the holographic 
one-point function $\langle \op(\vect{k}) \rangle_{\hat{\phi}}$
in terms of the response $\bar{\phi}$ 
and verified that after renormalization 
the two are directly proportional to one another, 
modulo a boundary term. Clearly, 
the specific functional form of the cubic interaction Lagrangian, 
$\delta L_3/\delta \phi$, and the boundary term, $\xi_2$, 
will depend on the nature of the effective field theory in
question. However, before we proceed to discuss the effective 
action for perturbations in the inflaton field,  
we will first discuss an alternative parametrization 
for the one-point function that will prove useful in what follows. 

\subsection{The Holographic One-point Function in terms of the Interaction 
Lagrangian}

\label{sec:4.2}

Following \cite{bianchi-prisco-muck}, we may quite generally 
define an operator, $\X$, such that 
\begin{equation}
  \label{threept:interim}
  \widetilde{\frac{\delta L_3}{\delta \phi(\eta,\vect{k_1})}}
  = \int \d^3 x \; \e{\imag \vect{k}_1 \cdot
  \vect{x}} \left[ \X(\vect{k}_1,\partial_2,\partial_3) \phi_0(\eta,\vect{x}_2)
  \phi_0(\eta,\vect{x}_3) \right]_{\vect{x}_2 = \vect{x}_3 = \vect{x}},
\end{equation}
where $\X(x,y,z)$ is a sum of powers of $x$, $y$ and $z$
and the derivatives $\partial_1$ and $\partial_2$ act 
only on $\vect{x}_1$ and $\vect{x}_2$, respectively. The functional 
form of $\X$ is determined by the nature of the cubic 
couplings in the interaction potential. 
We allow $\X$ to contain a $\vect{k}_1$ dependence
(where the subscript `$1$' is introduced for future convenience) in order
to accommodate interactions\footnote{The operator 
$\partial^{-2}$ is the solution operator for the Laplacian, 
defined such that $\partial^{-2} (\partial^2 \phi ) =\phi$.} of the form
$L_3 \propto \phi \partial^{-2} \phi \partial^2 \phi$ 
which, when written in terms of the field equations, generates a source
that includes terms of the form
$\delta L_3/\delta \phi \propto \partial^2 (\phi \partial^{-2} \phi)$.
For such sources, $\X$ can be written 
as the $k_1$ dependent expression $\X \propto -k_1^2 \partial_3^{-2}$
(for example).

When $L_3$ represents a cubic interaction,  
standard manipulations imply that Eq. \eref{threept:interim} can be written
as a convolution over two copies of $\phi_0$ such that 
\begin{eqnarray}
\fl\nonumber
  \widetilde{\frac{\delta L_3}{\delta \phi(\eta,\vect{k}_1)}}
  = \int \frac{\d^3 k_2 \,
  \d^3 k_3}{(2\pi)^3} \; \delta(\vect{k}_1+\vect{k}_2+\vect{k}_3)
  \X(\vect{k}_1,-\vect{k}_2,-\vect{k}_3)
  \phi_0(\eta,-\vect{k}_2) \phi_0(\eta,-\vect{k}_3) \\
  \label{threept:interima}
  \lo{=} \int \frac{\d^3 k_2 \, \d^3 k_3}{(2\pi)^3} \;
  \delta(\sum_i \vect{k}_i)
  \X_{123} K(\eta,k_2) K(\eta,k_3)
  \hat{\phi}(-\vect{k}_2) \hat{\phi}(-\vect{k}_3) ,
\end{eqnarray}
where in the second expression 
we have written $\phi_0$ in terms of the
bulk-to-boundary propagator \eref{twopt:soln}
and $\X_{123}$ is a convenient shorthand for 
$\X_{123} = \X(\vect{k}_1,-\vect{k}_2,-\vect{k}_3)$.
This is sufficient to parametrize the bare response, Eq. \eref{onept:phibar}, 
in the form 
\begin{eqnarray}
  \fl  \bar{\phi}^\ren(\vect{k}_1) =
  \imag \Im \Bigg(
  \frac{\imag k_1^3}{3} \hat{\phi}(\vect{k}_1) \nonumber \\
  \mbox{} +
  \frac{H^2}{3} \int \d\tau \int \frac{\d^3 k_2 \, \d^3 k_3}{(2\pi)^3}
  \; \delta(\sum_i \vect{k}_1) \X_{123} K_1 K_2 K_3 \hat{\phi}(-\vect{k}_2)
  \hat{\phi}(-\vect{k}_3) \Bigg) ,
\label{threept:resp}
\end{eqnarray}
where $K_j \equiv K(\tau,k_j)$. 

A similar parametrization may be employed for the (bare) boundary term,  
$\xi_2$. Indeed, by following an identical argument to that which led to
Eq. \eref{threept:interima}, we may write
\begin{equation}
  \fl\label{threept:xi}
  \xi_2^\ren(\eta,\vect{k}_1) = \imag \Im \left( \lim_{\eta \rightarrow 0}
  \int \frac{\d^3 k_2 \, \d^3 k_3}{(2\pi)^3}
  \; \delta(\sum_i \vect{k}_i) \Y_{123} K(\eta,k_2) K(\eta,k_3)
  \hat{\phi}(-\vect{k}_2) \hat{\phi}(-\vect{k}_3) \right) 
\end{equation}
and Eq. \eref{threept:xi} should be viewed as the definition of 
the quantity $\Y_{123} \equiv \Y (\vect{k}_1,-\vect{k}_2,-\vect{k}_3)$. 

Hence, after substituting Eqs. \eref{threept:resp} and \eref{threept:xi}
into Eq. \eref{dscft:onept} for the one-point function $\langle \op(\vect{k})
\rangle_{\hat{\phi}}$, we find that
\begin{eqnarray}
\fl\nonumber
\langle \op(\vect{k}_1) \rangle_{\hat{\phi}} 
  = \Re \Bigg( - \frac{1}{(2\pi )^3} \frac{k_1^3}{H^2} \hat{\phi} (-\vect{k}_1 )
  \\ \nonumber \hspace{-1cm} \mbox{}
  + \frac{\imag}{(2\pi)^3} \int_{-\infty}^{0}
  \d\tau \int \frac{\d^3 k_2\,\d^3 k_3}
  {(2\pi)^3} \delta(-\vect{k}_1 + \vect{k}_2 + \vect{k}_3) \X_{123}
  K_1 K_2 K_3 \hat{\phi}(-\vect{k}_2) \hat{\phi}(-\vect{k}_3) \\
  \hspace{-1cm} \label{onepointXY} \mbox{}
  + \frac{\imag}{(2\pi)^3} \lim_{\eta \rightarrow 0}
  \int \frac{\d^3 k_2 \, \d^3 k_3}
  {(2\pi)^3} \delta(-\vect{k}_1 + \vect{k}_2 + \vect{k}_3) \Y_{123}
  K_1 K_2 K_3 \hat{\phi}(-\vect{k}_2) \hat{\phi}(-\vect{k}_3) \Bigg) .
\end{eqnarray}

Now that the one-point function in the presence of the
source $\hat{\phi}$ has been determined, all higher
$n$-point functions may be derived from 
this quantity by functionally  differentiating with respect to
$\hat{\phi}$ and specifying $\hat{\phi} =0$ afterwards. 
In the following Section, we present the general 
expressions for the two- and three-point functions. 

\section{Holographic Two- and Three-point Functions}

\label{sec:5}

\subsection{Two-point Function}

\label{sec:5.1}

To obtain the two-point function, we substitute Eq. 
\eref{onept:phibar} into Eq. \eref{dscft:onept} and
differentiate. It follows that \cite{witten-ads}
\begin{equation}
  \label{twopt:value}
  \langle \op(\vect{k}_1) \op(\vect{k}_2) \rangle
  = \frac{\delta}{\delta \hat{\phi}(\vect{k}_2)}
  \left.\langle \op(\vect{k}_1) \rangle_{\hat{\phi}} \right|_{\hat{\phi}=0}
  = - \frac{1}{(2\pi)^3}\frac{k_1^3}{H^2} \delta(\vect{k}_1 + \vect{k}_2) .
\end{equation}
Note that there is no contribution to the two-point function from the surface
term $\xi_2$.

\subsection{Three-point Function}

\label{sec:5.2}

The three-point function is obtained from the second 
functional derivative of Eq. \eref{onepointXY}:  
\begin{equation}
  \label{threepointXY}
  \langle \op(\vect{k}_1) \op(\vect{k}_2) \op(\vect{k}_3) \rangle
  = \left. \frac{\delta}{\delta \hat{\phi}(\vect{k}_2)}
    \frac{\delta}{\delta \hat{\phi}(\vect{k}_3)}
    \langle \op(\vect{k}_1) \rangle_{\hat{\phi}} \right|_{\hat{\phi}=0} .
\end{equation}
The presence of two functional derivatives means that even though
we calculated to $\Or(g)$ in the term that is quadratic in $\hat{\phi}$,
it was not necessary to account for
$\Or(g)$ terms in the piece that is linear in $\hat{\phi}$. 
(This would have been
necessary to self-consistently solve the field equations to $\Or(g)$.)
Such terms would
correspond to  loop corrections to the two-point function of the sort
calulated in Ref. \cite{weinberg-corrl}. Hence, 
Eq.~\eref{threepointXY} yields the concise expression 
\begin{eqnarray}
  \fl\nonumber
  \label{concise}
  \langle \op(\vect{k}_1) \op (\vect{k}_2) \op (\vect{k}_3) \rangle
  = \Re \frac{2 \imag}{(2\pi)^6} \delta(\sum_i \vect{k}_i) \Bigg[
  \int^0_{-\infty} \d \tau \;
  {\symm}_{1, 2, 3} \left( \X_{123} K_1 K_2 K_3
  \right) \\ \label{threept:expr}
  \mbox{} + \lim_{\eta \rightarrow 0}
  \int \; {\symm}_{1, 2, 3} \left( \Y_{123} K_1 K_2 K_3
  \right) \Bigg] ,
\end{eqnarray}
where $\symm_{1, 2, 3}$ denotes symmetrization with weight unity over the
labels $1$,
$2$ and $3$. This symmetrization must occur since Eq. \eref{threept:expr}
was derived from Eq. \eref{dscft:cut-offs} by three functional variations.
In Ref. \cite{bianchi-prisco-muck}, this symmetrization was
achieved in a different way, by symmetrization over $2 \leftrightarrow 3$
and integrating the result by parts to obtain symmetry over the labels
$1$, $2$ and $3$ in combination. However, the result must be the 
same.\footnote{There is a potential subtlety with this argument, because the
one-point function is calculated from the regularized action, with a cut-off
at $\eta = -\epsilon$. The cut-off is then removed, and higher correlation
functions are obtained by functional differentiation. Thus, the two- and
three-point functions are not precisely obtained by functional variation of
a bosonic action, because of the intervening limiting process.
This point was made explicitly in
\cite{bianchi-prisco-muck}. We assume this subtlety makes no difference to
the correlation functions we are calculating.}
Note also that the combination of signs for the various momenta $\vect{k}_i$
which appear in Eq. \eref{onepointXY} is necessary to obtain the
correct expression for momentum conservation in Eq. \eref{threept:expr}.
We have assumed that $\X_{123}$ and $\Y_{123}$ do not change under a change
of sign $\vect{k}_i \mapsto -\vect{k}_i$. It is easy to verify that this is
the case for the effective field theory of inflation, because $\X_{123}$ and
$\Y_{123}$ depend on momentum only via the combinations $\partial^2$ and
$\partial^{-2}$, which are invariant under sign exchange. 
In more general theories, however, 
sign changes in $\vect{k}_i$ may introduce extra signs in \eref{threept:expr},
which should be correctly accounted for.

Thus far, the discussion has been entirely general,  
in the sense that we have not yet specified the precise 
form of the cubic interactions. 
In the following Section, we proceed to consider the case that is relevant to 
inflationary perturbation theory. 

\section{Effective Field Theory for the Inflaton}

\label{sec:6}

\subsection{Third-order Action}

\label{sec:6.1}

We consider a universe sourced by a single 
scalar `inflaton' field $\varphi$ that is 
minimally coupled to Einstein gravity and self-interacting through a 
potential $W(\varphi)$. The four-dimensional bulk action is given by 
\begin{equation}
  \label{dscft:sugra}
  S =
  \int \d^4 x \; \sqrt{-\det g}
  \left[ \frac{1}{2} R - \frac{1}{2} (\partial \varphi)^2
  - W(\varphi) \right] .
\end{equation}

We assume that the background solution corresponds to 
a homogeneous scalar field $\varphi(\eta)$ propagating
on the de Sitter spacetime. The evolution of fluctuations 
in the inflaton field may then be described by introducing small
spatially dependent perturbations in the
metric and the scalar field and expanding the action \eref{dscft:sugra}
in powers of these perturbations. It proves most convenient to work 
in the uniform curvature gauge and we denote the (gauge-invariant) 
scalar field perturbation by $q$. This can be related via a change of gauge
to the comoving curvature perturbation $\R$ or the uniform-density curvature 
perturbation $\zeta$. The gravitational perturbations can then be directly 
related to $q$ via constraint equations.

Expanding action \eref{dscft:sugra} to third order in $q$ and to
leading order in the slow-roll parameter $\dot{\varphi}/H$ implies that 
the second- and third-order contributions to the action take the 
form \cite{seery-lidsey-a,maldacena-nongaussian}
\begin{eqnarray}
\label{secondorderaction}
  S_2 = \int \d \eta \, \d^3 x \; 
  \left[ \frac{1}{2} a^2 q^{\prime 2} - \frac{1}{2} a^2 (\partial q)^2 
\right] \\
\label{infl:action}
  S_3 = - \int \d \eta \, \d^3 x \;  \left[
  a q' \partial \psi \partial q + \frac{a^2\dot{\varphi}}{4H}
  q q^{\prime 2} + \frac{a^2 \dot{\varphi}}{4H} q (\partial q)^2
  \right] + \cdots  ,
\end{eqnarray}
where `$\cdots$' denote terms which contain higher powers
of $q$ or $\dot{\varphi}/H$,
and, as discussed after Eq.~\eref{threepointXY}, we include only the leading
order slow-roll piece at each order in $q$. The auxiliary function $\psi$ is 
a gravitational perturbation and satisfies the constraint equation
\begin{equation}
  \partial^2 \psi = - \frac{a \dot{\varphi}}{2H} q' .
\end{equation}
Eq.~\eref{infl:action} follows directly from 
Eq.~\eref{dscft:sugra} without performing any field redefinitions or 
integrations by parts in $\eta$, which would produce auxiliary boundary
terms at $\eta = 0^-$. On the other hand we have freely integrated by
parts in the spatial variables $x^i$ which do not produce any surface terms.

After substituting for the gravitational 
perturbation, $\psi$, one finds that the cubic interaction terms
in Eq. \eref{infl:action} can be rewritten in the form
(see also \cite{balasubramanian-boer-b})
\begin{eqnarray}
  \fl\nonumber
  S_3 = \int \d \eta \, \d^3 x \; \left[
  - \frac{\dot{\varphi}}{4H} \partial^{-2} \left. \frac{\delta L}{\delta q}
  \right|_1 (\partial q)^2 - \frac{a^2 \dot{\varphi}}{2H} q' \partial^{-2}
  q' \partial^2 q - \frac{a^2\dot{\varphi}}{4H} q q^{\prime 2}
  \right] \\
  \label{infl:boundary}
  \mbox{} - \int_{\eta = 0} \d^3 x \; \frac{a^2\dot{\varphi}}{4H}
  \partial^{-2} q' (\partial q)^2 ,
\end{eqnarray}
where $\delta L/\delta q|_1$ is the first-order equation of motion:
\begin{equation}
  \frac{1}{a^2} \left. \frac{\delta L}{\delta q} \right|_1 =
  \frac{2}{\eta} q' - q'' + \partial^2  q .
\end{equation}
Thus, $\delta L/\delta q|_1 = 0$ when evaluated on the classical solution.

The boundary term in Eq. \eref{infl:boundary} may be removed 
by performing a field redefinition of the form $q \mapsto \phi + F(q)$. 
This implies that the second-order action \eref{secondorderaction}
transforms to 
\begin{equation}
  S_2[q] \mapsto S_2[\phi] 
  + \int \d\eta \, \d^3 x \; F(q) \left. \frac{\delta L}{\delta q}
  \right|_1 + \int_{\eta=0} \d^3 x \; a^2 q' F(q) .
\end{equation}
Hence, specifying $F$ so that it satisfies the condition 
\begin{equation}
\label{defF}
  F = \frac{\dot{\varphi}}{4H} \partial^{-2} (\partial q)^2
\end{equation}
implies that the field redefinition $q \mapsto \phi + F(q)$ 
removes both the $\delta L/
\delta q|_1$ term and the boundary integral $\int_\partial$ 
from the third-order action \eref{infl:boundary}.  
A reduced interaction term of the form
\begin{equation}
  \label{infl:phi}
  S_3 = \int \d \eta \, \d^3 x \; \left[ - \frac{a^2\dot{\varphi}}{2H}
  \phi' \partial^{-2} \phi' \partial^2 \phi -
  \frac{a^2\dot{\varphi}}{4H} \phi \phi'^2   \right] 
\end{equation}
is all that remains. 

\subsection{Holographic Three-point Function for the Inflaton}

\label{sec:6.2}

We may now derive the (renormalized) holographic three-point function
of the deformed  Euclidean CFT that is dual to the inflaton. 
We begin by varying the interaction sector of the bulk action, 
Eq. \eref{infl:phi}, in order to determine the specific forms of the 
contributions $\X_{123}$ and $\Y_{123}$ that arise in the general 
expression for the three-point
correlation function, Eq.  \eref{concise}. Variation of the action 
\eref{infl:phi} results in both bulk and boundary contributions,  
as summarized in Eq. \eref{onept:deltaLrule}. After comparing the 
former to the general expression \eref{threept:interima}, 
we deduce an expression for $\X_{123}$ of the form 
\begin{eqnarray}
  \nonumber\fl
  \X_{123} = \frac{a^2 \dot{\varphi}}{2H} \Bigg[
  \left(\frac{1}{2} + \frac{k_3^2}{k_1^2} + \frac{k_3^2}{k_2^2} -
  \frac{k_1^2}{k_2^2} \right) \partial_2 \partial_3 +
  \left( \frac{k_3^2}{k_2^2} + \frac{k_3^2}{k_1^2} \right) \partial_2^2 +
  \partial_3^2 \\\label{infl:x}
  \lo{+} 2 \frac{a'}{a} \left( \frac{k_3^2}{k_2^2} + \frac{k_3^2}{k_1^2}
  \right) \partial_2 + 2 \frac{a'}{a} \partial_3 \Bigg] ,
\end{eqnarray}
where $\partial_n$ denotes an $\eta$-derivative which acts in 
Eq. \eref{threept:interima} on the
bulk-to-boundary propagator $K_n$, where 
$K_n$ is given by Eq. \eref{twopt:green}. 
 
Likewise, the form of the surface term is deduced
by comparing the boundary terms that arise in the variation of the 
action \eref{infl:phi} directly with Eq. \eref{threept:xi}. 
We find that 
\begin{equation}
  \label{infl:y}
  \Y_{123} = - \frac{a^2 \dot{\varphi}}{2H} \left[
  \left(\frac{k_3^2}{k_2^2} + \frac{k_3^2}{k_1^2}\right) \partial_2 +
  \partial_3 \right] .
\end{equation}

It only remains to evaluate the integrals in the 
three-point function \eref{threept:expr}. Let us consider the 
first integral in this expression. For a 
bulk-to-boundary propagator of the form \eref{twopt:green}, 
we find after integration by parts that
\begin{equation}
  \label{second:interim}
  \int_{-\infty}^0 \d \tau \; \X_{123} K_1 K_2 K_3 =
  \int_{-\infty}^0 \d \tau \; \bar{\X}_{123} K_1 K_2 K_3 -
  \int_{\eta=0} \Y_{123} K_1 K_2 K_3 ,
\end{equation}
where $\bar{\X}$ is defined by 
\begin{equation}
  \bar{\X}_{123} = \frac{a^2 \dot{\varphi}}{2H} \left[
  \left( - \frac{1}{2} - \frac{k_1^2}{k_2^2}
  \right) \partial_2 \partial_3 - \partial_1 \partial_3 -
  \left( \frac{k_3^2}{k_2^2} + \frac{k_3^2}{k_1^2} \right)
  \partial_1 \partial_2 \right] .
\end{equation}
Hence, the boundary term $\Y_{123}$ in the three-point correlator 
\eref{threept:expr} is {\em precisely canceled} by the second term 
in Eq. \eref{second:interim}. 
This implies that the three-point correlator may be simplified to 
\begin{equation}
  \fl\nonumber
   \label{oneintegral}
  \langle \op(\vect{k}_1) \op (\vect{k}_2) \op (\vect{k}_3) \rangle
  = \Re \frac{2 \imag}{(2\pi)^6} \delta(\sum_i \vect{k}_i) 
  \int^0_{-\infty} \d \tau \;
  {\symm}_{1, 2, 3} \left( \bar{\X}_{123} K_1 K_2 K_3
  \right)  .
\end{equation}

In order to evaluate the integral in Eq. \eref{oneintegral},  
we employ the result: 
\begin{equation}
  \label{second:a}
  \imag \int_{-\infty}^0 \frac{\d \eta}{\eta^2} \; \partial_2 \partial_3
  K_1 K_2 K_3 = \frac{k_2^2 k_3^2}{k_t} + \frac{k_1 k_2^2 k_3^2}{k_t^2} ,
\end{equation}
where Eq. \eref{twopt:green} has been employed once more
and $k_t \equiv k_1+k_2+k_3$. It follows, therefore, that
\begin{equation}
  \label{second:int}
  \imag \int_{-\infty}^0 \d \eta \; \symm_{1, 2, 3} \left[
  \bar{\X}_{123} K_1 K_2 K_3 \right] = - \frac{\dot{\varphi}}{H^3}
  \frac{1}{k_t} \sum_{i < j} k_i^2 k_j^2 
\end{equation}
and this implies that the three-point function may be 
expressed in the succinct form
\begin{equation}
\label{actualcalc}
\langle \op(\vect{k}_1) \op(\vect{k}_2) \op(\vect{k}_3) \rangle =
  - \frac{1}{(2\pi)^6} \delta(\sum_i \vect{k}_i) 
\frac{2 \dot{\varphi}}{H^3 k_t} \sum_{i < j} k_i^2 k_j^2 .
\end{equation}

\section{The Inflaton Three-point Function}

\label{sec:7}

The dictionary that relates the correlators for the bulk inflaton field
to those for the dual CFT as calculated
above was presented in Refs. \cite{maldacena-nongaussian,schaar}. 
In particular, the two-point functions are related by 
\begin{equation}
  \label{corrl:twopt}
  \langle \phi(\vect{k}_1) \phi(\vect{k}_2) \rangle =
  - \frac{1}{2 \Re \langle \op(\vect{k}_1) \op (\vect{k}_2) \rangle} ,
\end{equation}
and the analogous expression for the three-point functions 
is given by 
\begin{equation}
  \label{corrl:threept}
  \langle \phi(\vect{k}_1) \phi(\vect{k}_2) \phi(\vect{k}_3) \rangle
  = \frac{2 \Re \langle \op(\vect{k}_1) \op(\vect{k}_2) \op(\vect{k}_3)
  \rangle}{\prod_i (-2 \Re \langle \op(\vect{k}_i) \op(-\vect{k}_i) 
\rangle')} ,
\end{equation}
where $\langle \op \op \rangle'$ represents the correlator with the 
momentum-conservation $\delta$-function omitted.

Comparison between Eqs. \eref{twopt:value} and \eref{corrl:twopt}
immediately yields the two--point function for the inflaton field 
\cite{schaar}: 
\begin{equation}
\label{bulk2point}
\langle \phi(\vect{k}_1) \phi(\vect{k}_2) \rangle =
(2 \pi )^3 \frac{H^2}{2k_1^3} \delta ( \vect{k}_1+\vect{k}_2 )  .
\end{equation}

The three-point function follows after substitution of 
Eqs. \eref{actualcalc} and \eref{bulk2point} into Eq. \eref{corrl:threept}: 
\begin{equation}
  \label{second:result}
  \langle \phi(\vect{k}_1) \phi(\vect{k}_2) \phi(\vect{k}_3) \rangle =
  - (2\pi)^3 \delta(\sum_i \vect{k}_i) \frac{4\dot{\varphi} H^3}{\prod_i
  2k_i^3} \frac{1}{k_t} \sum_{i < j} k_i^2 k_j^2 .
\end{equation}

The corresponding three-point function for the inflaton field perturbation, 
$q$, can then be determined by introducing the field 
redefinition (\ref{defF}) back into Eq. \eref{second:result} and 
using Wick's theorem, 
as described in \cite{maldacena-nongaussian,seery-lidsey-a}. 
This yields the final result:  
\begin{eqnarray}
  \label{finalresult}
  \fl
  \langle q(\vect{k}_1) q(\vect{k}_2) q(\vect{k}_3) \rangle =
  (2\pi)^3 \delta(\sum_i \vect{k}_i) \frac{1}{\prod_i
  2k_i^3} \dot{\varphi} H^3 \nonumber \\
  \mbox{} \times \left[ \frac{1}{2} \sum_i k_i^3 
  - \frac{4}{k_t} \sum_{i<j} k_i^2k_j^2
  - \frac{1}{2} \sum_{i\ne j} k_ik_j^2 \right] .
\end{eqnarray}
Eq. \eref{finalresult} is expression (68) of Ref. \cite{seery-lidsey-a} 
specialized to a single field, which was derived from the
bulk QFT calculation. 

However, it is not necessary to perform a field redefinition 
in order to take into account the boundary terms 
of the form $\int_{\eta=0}$ in the third-order action \eref{infl:boundary}. 
The three-point correlator \eref{finalresult} may also be 
derived by including the boundary term explicitly. The form of this  
term is given by 
\begin{equation}
\label{boundary}
  \mbox{} - \int_{\eta=0} \d^3 x \; \frac{a^2\dot{\varphi}}{4H}
  \partial^{-2} q' (\partial q)^2 .
\end{equation}
This is manifestly divergent at
future infinity, because the scale factor $a$ is unbounded at late
times. This divergence is subtracted using our renormalization prescription,
Eq.~\eref{boundary}, and one may verify that it is purely imaginary.
Furthermore, the oscillatory nature of the wavefunction
at past infinity implies that there will be 
no contribution from regions where
a given $k$-mode is deep inside the horizon \cite{maldacena-nongaussian,
weinberg-corrl}. 

The boundary integral \eref{boundary}
may be evaluated when the Bunch-Davies vacuum 
is invoked as the initial state for the perturbation. In this 
case, the perturbation evolves as 
\begin{equation}
\label{formq}
  q = \frac{(2\pi )^{3/2}}{\sqrt{2k^3}} H  (1-\imag k\eta ) e^{ik \eta}  ,
\end{equation}
which implies that, after renormalization, the
boundary term yields a contribution 
to the $q$-correlator of the form
\begin{equation}
  \fl\label{infl:divergent}
  \Delta \langle q(\vect{k}_1) q(\vect{k}_2) q(\vect{k}_3) \rangle =
  (2\pi)^3 \delta ( \sum_i \vect{k}_i ) \frac{H^3 \dot{\varphi}}
  {\prod_i 2k_i^3} \frac{\vect{k}_2 \cdot \vect{k}_3}{4} k_1 + 
  \mbox{perms} + \mbox{c.c.} ,
\end{equation}
where c.c. denotes the complex conjugate. 
Finally, after taking the permutations into account and using the relation 
$\sum_i \vect{k}_i = 0$, we find that the contribution 
of the boundary term is 
\begin{equation}
  \label{infl:qboundary}
  \fl
  \Delta \langle q(\vect{k}_1) q(\vect{k}_2) q(\vect{k}_3) \rangle =
  (2\pi)^3 \delta ( \sum_i \vect{k}_i ) \frac{4\pi^4}{\prod_i k_i^3}
  \left( \frac{H}{2\pi} \right)^4
  \frac{\dot{\varphi}}{4H} \left( \sum_i k_i^3 - \sum_{i \neq j} k_i k_j^2
  \right) .
\end{equation}
Combining Eqs. \eref{second:result} and \eref{infl:qboundary} therefore
exactly reproduces the three-point correlator for the 
inflaton perturbation, Eq. \eref{finalresult}. 

\section{Discussion}

\label{sec:8}

In this paper, we have demonstrated explicitly how the formalism of
the dS/CFT correspondence may be employed to derive the primordial 
three-point correlation function of the inflaton field perturbation
by calculating the bulk prediction for the
corresponding three-point CFT correlator. This complements the standard
bulk QFT approach based on evaluating tree-level Feynman diagrams.
It also provides an important consistency check of 
Maldacena's formula \cite{maldacena-nongaussian,schaar} relating
the bulk and boundary correlators.
We have also emphasized that the inflaton three-point function
can be determined directly from the third-order action \eref{infl:boundary}
without the need for a field redefinition if the contribution from  
the boundary integral is included. Although, for simplicity, we have
limited the analysis to a single field, the extension to multiple-field
models is straightforward.

When derivative interactions are present in the action, boundary terms at
future infinity arise after varying the action to obtain the field equations.
These terms have not previously been considered explicitly in the AdS/CFT
case. In the analysis of Ref. \cite{muck-viswanathan}, such interactions
may implicitly be present, but the associated boundary terms are included
in the field equations or are taken to vanish,
which could be achieved either by picking appropriate boundary conditions
for the fields, or by modifying the action. As is usual in de Sitter field
theory, we have discarded the contribution of these terms from the bulk field
equation. This has important 
consequences for the one-point correlation function of the dual CFT. 
We have shown explicitly in Eq. \eref{dscft:onept} that 
the one-point function in the dS/CFT correspondence is indeed proportional 
to the response, as given by Eq. \eref{onept:phibar}, but only 
modulo a boundary term. This term contributes  
to the three-point function, as seen in Eq. \eref{concise}.   
However, such a contribution is cancelled after the term that 
contains the variation of the cubic interaction Lagrangian is 
integrated by parts to remove higher-derivative operators.
Hence, the three-point function 
can be calculated even if the form of the boundary contribution is unknown. 

One of the attractive features of the method we have outlined above is that 
the holographic one-point function is determined by the variation of the 
interaction Lagrangian. This suggests that it should be possible in principle 
to determine the correlator from the {\em second-order} field equation
and, furthermore, indicates that the
$n$-point function of the primordial inflaton perturbation could be 
determined by employing only $(n-1)^{\mathrm{th}}$-order perturbation 
theory. This is in contrast to conventional approaches, where 
the action must be evaluated to $\Or(q^n)$ in order to obtain the
$n$-point $q$-correlator. 

To calculate the $n$-point function in this way, one would begin
with the full action for Einstein gravity coupled to a scalar field
(c.f. \cite{malik}),
$\varphi$, which can be written in the form \cite{seery-lidsey-a}
\begin{equation}
  \fl
  \label{totalaction}
  S = - \frac{1}{2} \int N \sqrt{h} \left[ \grad^i \varphi \grad_i \varphi +
  2 W(\varphi) \right] + \frac{1}{2} \int \frac{\sqrt{h}}{N} \left[
  E^{ij} E_{ij} - E^2 + \pi^2 \right] ,
\end{equation}
where we have adopted the ADM line element,
\begin{equation}
  \d s^2 = - N^2 \d t^2 + h_{ij} (\d x^i + N^i)(\d x^j + N^j)
\end{equation}
with lapse function $N$ and shift vector $N^j$, and
$\pi$ is the scalar field momentum, $\pi = \dot{\varphi} - N^j \grad_j
\varphi$. The spatial tensor $E_{ij}$ is related to the extrinsic curvature
of the spatial slices such that 
$E_{ij} = \frac{1}{2} \dot{h}_{ij} - \grad_{(i}N_{j)}$.
Eq.~\eref{totalaction} should be evaluated by selecting a spatial gauge
for $h_{ij}$, such as the comoving gauge where $h_{ij} = a^2(t) \delta_{ij}$,
and parametrizing the inflaton field in terms of a homogeneous background
component, $\varphi_h$, and a perturbation $q$, i.e., 
$\varphi =\varphi_h +q$. 

The equations of motion for the fields $N$ and $N_j$ are given by 
\begin{equation}
  - \grad^i \varphi \grad_i \varphi - 2W(\varphi) - 
\frac{1}{N^2} (E^{ij}E_{ij} - E^2
  + \pi^2) = 0
\end{equation}
and
\begin{equation}
  \grad_j \left( \frac{1}{N} [ E^j_i - E \delta^j_i ] \right) = \frac{1}{N}
  \pi \grad_i \varphi ,
\end{equation}
respectively, and their solution specifies $N = N(\varphi_h,q)$ and 
$N_j = N_j(\varphi_h,q)$ as
functions of the background inflaton trajectory and the perturbation $q$.
The field equation for the inflaton fluctuation is also 
required and this follows by varying the action \eref{totalaction}
with respect to $q$. 
This field equation can then be used to determine the quantities
$\X_{123}$ and $\Y_{123}$, from which 
the renormalized response $\bar{\phi}^\ren$
and boundary term $\xi_2^\ren$ follow, thereby yielding the 
correlator.  

In general, however, the obstacle to this procedure is that the solution
of the constraint equations for the lapse and shift  
is a complicated function of the inflaton perturbation $q$. Consequently, 
when determining the three-point correlator, it is as straightforward 
to reduce the action \eref{totalaction} 
to a functional in $q^3$ and calculate the 
correlator using Feynman graphs, as it is to use the field 
equation in the holographic approach. On the other hand, for $4$- and higher
$n$-point functions, the prospect of reducing the order of perturbation 
theory by one may provide a significant technical simplification. This 
will become more relevant in the future as the quality of CMB data 
improves. Indeed, constraints on inflationary non-Gaussianity
based on the primordial trispectrum have recently been discussed  
\cite{wmap2006-params,lyth-curvaton-ng,kogo-komatsu} and it is of 
importance to explore these issues further. 

\ackn
DS is supported by PPARC.

\providecommand{\href}[2]{#2}\begingroup\raggedright\endgroup


\begin{thebibliography}{10}

\bibitem{bartolo-matarrese-review}
N.~Bartolo, E.~Komatsu, S.~Matarrese, and A.~Riotto, {\it Non-{G}aussianity
  from {I}nflation: {T}heory and {O}bservations},  {\em Phys. Rept.} {\bf 402}
  (2004) 103--266, [\href{http://xxx.lanl.gov/abs/astro-ph/0406398}{{\tt
  astro-ph/0406398}}].

\bibitem{maldacena-nongaussian}
J.~Maldacena, {\it Non-{G}aussian features of primordial fluctuations in single
  field inflationary models},  {\em JHEP} {\bf 0305} (2003) 013,
  [\href{http://xxx.lanl.gov/abs/astro-ph/0210603}{{\tt astro-ph/0210603}}].

\bibitem{seery-lidsey}
D.~Seery and J.~E. Lidsey, {\it Primordial non-gaussianity in single-field
  inflation},  {\em JCAP} {\bf 0506} (2005) 003,
  [\href{http://xxx.lanl.gov/abs/astro-ph/0503692}{{\tt astro-ph/0503692}}].

\bibitem{seery-lidsey-a}
D.~Seery and J.~E. Lidsey, {\it Primordial non-gaussianity from multiple-field
  inflation},  {\em JCAP} {\bf 0509} (2005) 011,
  [\href{http://xxx.lanl.gov/abs/astro-ph/0506056}{{\tt astro-ph/0506056}}].

\bibitem{creminelli}
P.~Creminelli, {\it On non-gaussianities in single-field inflation},  {\em
  JCAP} {\bf 0310} (2003) 003,
  [\href{http://xxx.lanl.gov/abs/astro-ph/0306122}{{\tt astro-ph/0306122}}].

\bibitem{lyth-rodriguez}
D.~Lyth and Y.~Rodr\'{\i}guez, {\it Non-gaussianity from the second-order
  cosmological perturbation},  {\em Phys. Rev. D} {\bf 71} (2005) 123508,
  [\href{http://xxx.lanl.gov/abs/astro-ph/0502578}{{\tt astro-ph/0502578}}].

\bibitem{lyth-rodriguez-a}
D.~Lyth and Y.~Rodr\'{\i}guez, {\it The inflationary prediction for primordial
  non-gaussianity},  {\em Phys. Rev. Lett.} {\bf 95} (2005) 121302,
  [\href{http://xxx.lanl.gov/abs/astro-ph/0504045}{{\tt astro-ph/0504045}}].

\bibitem{acquaviva-bartolo}
V.~Acquaviva, N.~Bartolo, S.~Matarrese, and A.~Riotto, {\it Second-{O}rder
  {C}osmological {P}erturbations from {I}nflation},  {\em Nucl. Phys. B} {\bf
  667} (2003) 119--148, [\href{http://xxx.lanl.gov/abs/astro-ph/0209156}{{\tt
  astro-ph/0209156}}].

\bibitem{lyth-zaballa}
D.~Lyth and I.~Zaballa, {\it A {B}ound {C}oncerning {P}rimordial
  {N}on-{G}aussianity},  {\em JCAP} {\bf 0510} (2005) 005,
  [\href{http://xxx.lanl.gov/abs/astro-ph/0507608}{{\tt astro-ph/0507608}}].

\bibitem{allen-gupta}
L.~Allen, S.~Gupta, and D.~Wands, {\it Non-{G}aussian perturbations from
  multi-field inflation},  \href{http://xxx.lanl.gov/abs/astro-ph/0509719}{{\tt
  astro-ph/0509719}}.

\bibitem{zaballa-rodriguez}
I.~Zaballa, Y.~Rodr\'{\i}guez, and D.~H. Lyth, {\it {Higher order contributions
  to the primordial non-gaussianity}},
  \href{http://xxx.lanl.gov/abs/astro-ph/0603534}{{\tt astro-ph/0603534}}.

\bibitem{malik}
K.~Malik, {\it {G}auge-invariant perturbations at second order: multiple scalar
  fields on large scales},  {\em JCAP} {\bf 0511} (2005) 005,
  [\href{http://xxx.lanl.gov/abs/astro-ph/05060532}{{\tt astro-ph/05060532}}].

\bibitem{rigopoulos-shellard}
G.~Rigopoulos and E.~Shellard, {\it Non-linear inflationary perturbations},
  {\em JCAP} {\bf 0510} (2005) 006,
  [\href{http://xxx.lanl.gov/abs/astro-ph/0405185}{{\tt astro-ph/0405185}}].

\bibitem{rigopoulos-shellard-vantent}
G.~Rigopoulos, E.~Shellard, and B.~van Tent, {\it Non-linear perturbations in
  multiple-field inflation},
  \href{http://xxx.lanl.gov/abs/astro-ph/0504508}{{\tt astro-ph/0504508}}.

\bibitem{calcagni-nongaussian}
G.~Calcagni, {\it Non-{G}aussianity in braneworld and tachyon inflation},  {\em
  JCAP} {\bf 0510} (2005) 009,
  [\href{http://xxx.lanl.gov/abs/astro-ph/0411773}{{\tt astro-ph/0411773}}].

\bibitem{alabidi-lyth}
L.~Alabidi and D.~Lyth, {\it Inflation models and observation},
  \href{http://xxx.lanl.gov/abs/astro-ph/0510443}{{\tt astro-ph/0510443}}.

\bibitem{enqvist-jokinen}
K.~Enqvist, A.~Jokinen, A.~Mazumdar, T.~Multamaki, and A.~V\"{a}ihk\"{o}nen,
  {\it Non-{G}aussianity from {P}reheating},  {\em Phys. Rev. Lett.} {\bf 94}
  (2005) 161301, [\href{http://xxx.lanl.gov/abs/astro-ph/0411394}{{\tt
  astro-ph/0411394}}].

\bibitem{jokinen-mazumdar}
A.~Jokinen and A.~Mazumdar, {\it Very {L}arge {P}rimordial {N}on-{G}aussianity
  from multi-field: {A}pplication to {M}assless {P}reheating},
  \href{http://xxx.lanl.gov/abs/astro-ph/0512368}{{\tt astro-ph/0512368}}.

\bibitem{barnaby-cline}
N.~Barnaby and J.~M. Cline, {\it Nongaussian and nonscale-invariant
  perturbations from tachyonic preheating in hybrid inflation},
  \href{http://xxx.lanl.gov/abs/astro-ph/0601481}{{\tt astro-ph/0601481}}.

\bibitem{wmap2006-params}
D.~Spergel, R.~Bean, O.~Dor\'e, M.~Nolta, C.~Bennett, G.~Hinshaw, N.~Jarosik,
  E.~Komatsu, L.~Page, H.~Peiris, L.~Verde, C.~Barnes, M.~Halpern, R.~Hill,
  A.~Kogut, M.~Limon, S.~Meyer, N.~Odegard, G.~Tucket, J.~Weiland, E.~Wollack,
  and E.~Wright, {\it Wilkinson {M}icrowave {A}nisotropy {P}robe ({WMAP})
  {T}hree {Y}ear {R}esults: {I}mplications for {C}osmology},
  \href{http://xxx.lanl.gov/abs/astro-ph/0603449}{{\tt astro-ph/0603449}}.

\bibitem{wmap2006-temp}
G.~Hinsaw, M.~Nolta, C.~Bennett, R.~Bean, O.~Dor\'e, M.~Greason, M.~Halpern,
  R.~Hill, N.~Jarosik, A.~Kogut, E.~Komatsu, M.~Limon, N.~Odegard, S.~Meyer,
  L.~Page, H.~Peiris, D.~Spergel, G.~Tucker, L.~Verde, J.~Weiland, E.~Wollack,
  and E.~Wright, {\it Three-{Y}ear {W}ilkinson {M}icrowave {A}nisotropy {P}robe
  ({WMAP}) {O}bservations: {T}emperature {A}nalysis},
  \href{http://xxx.lanl.gov/abs/astro-ph/0603451}{{\tt astro-ph/0603451}}.

\bibitem{mcewen-hobson}
J.~McEwen, M.~Hobson, A.~Lasenby, and D.~Mortlock, {\it A high-significance
  detection of non-{G}aussianity in the {WMAP} 1-year data using directional
  spherical wavelets},  {\em Mon. Not. Roy. Astron. Soc.} {\bf 359} (2005)
  1583--1596, [\href{http://xxx.lanl.gov/abs/astro-ph/0406604}{{\tt
  astro-ph/0406604}}].

\bibitem{mukherjee-wang}
P.~Mukherjee and Y.~Wang, {\it Wavelets and {WMAP} non-{G}aussianity},
  \href{http://xxx.lanl.gov/abs/astro-ph/0402602}{{\tt astro-ph/0402602}}.

\bibitem{vielva-martinez-gonzalez}
P.~Vielva, E.~Martinez-Gonzalez, R.~Barreiro, J.~Sanz, and L.~Cayon, {\it
  Detection of non-{G}aussianity in the {WMAP} 1-year data using spherical
  wavelets},  {\em Astrophys. J.} {\bf 609} (2004) 22--34,
  [\href{http://xxx.lanl.gov/abs/astro-ph/0310273}{{\tt astro-ph/0310273}}].

\bibitem{larsen-wandelt}
D.~Larsen and B.~Wandelt, {\it A statistically robust 3-sigma detection of
  non-{G}aussianity in the {WMAP} data using hot and cold spots},
  \href{http://xxx.lanl.gov/abs/astro-ph/0505046}{{\tt astro-ph/0505046}}.

\bibitem{alishahiha-silverstein}
M.~Alishahiha, E.~Silverstein, and D.~Tong, {\it D{BI} in the {S}ky},  {\em
  Phys. Rev. D} {\bf 70} (2004) 123505,
  [\href{http://xxx.lanl.gov/abs/hep-th/0404084}{{\tt hep-th/0404084}}].

\bibitem{arkani-hamed-creminelli}
N.~Arkani-Hamed, P.~Creminelli, S.~Mukhoyama, and M.~Zaldarriaga, {\it Ghost
  {I}nflation},  {\em JCAP} {\bf 0404} (2004) 001,
  [\href{http://xxx.lanl.gov/abs/hep-th/0312100}{{\tt hep-th/0312100}}].

\bibitem{boubekeur-lyth}
L.~Boubekeur and D.~Lyth, {\it Detecting a small perturbation through its
  non-{G}aussianity},  \href{http://xxx.lanl.gov/abs/astro-ph/0504046}{{\tt
  astro-ph/0504046}}.

\bibitem{lyth-curvaton-ng}
D.~Lyth, {\it Non-gaussianity and cosmic uncertainty in curvaton-type models},
  \href{http://xxx.lanl.gov/abs/astro-ph/0602285}{{\tt astro-ph/0602285}}.

\bibitem{vernizzi-wands}
F.~Vernizzi and D.~Wands, {\it Non-{G}aussianities in two-field inflation},
  \href{http://xxx.lanl.gov/abs/astro-ph/0603799}{{\tt astro-ph/0603799}}.

\bibitem{weinberg-corrl}
S.~Weinberg, {\it {Q}uantum {C}ontributions to {C}osmological {C}orrelations},
  {\em Phys. Rev. D} {\bf 72} (2005) 043514,
  [\href{http://xxx.lanl.gov/abs/hep-th/0506236}{{\tt hep-th/0506236}}].

\bibitem{strominger}
A.~Strominger, {\it The d{S}/{CFT} {C}orrespondence},  {\em JHEP} {\bf 0110}
  (2001) 034, [\href{http://xxx.lanl.gov/abs/hep-th/0106113}{{\tt
  hep-th/0106113}}].

\bibitem{strominger-inflation}
A.~Strominger, {\it Inflation and the d{S}/{CFT} {C}orrespondence},  {\em JHEP}
  {\bf 0111} (2001) 049, [\href{http://xxx.lanl.gov/abs/hep-th/0110087}{{\tt
  hep-th/0110087}}].

\bibitem{klemm}
D.~Klemm, {\it Some {A}spects of the de {S}itter/{CFT} {C}orrespondence},  {\em
  Nucl. Phys. B} {\bf 625} (2002) 295--311,
  [\href{http://xxx.lanl.gov/abs/hep-th/0106247}{{\tt hep-th/0106247}}].

\bibitem{leblond-myers}
F.~Leblond, R.~Myers, and D.~Marolf, {\it Tall tales from de {S}itter space
  {I}: renormalization group flows},  {\em JHEP} {\bf 0206} (2002) 052,
  [\href{http://xxx.lanl.gov/abs/hep-th/0202094}{{\tt hep-th/0202094}}].

\bibitem{larsen-schaar}
F.~Larsen, J.~van~der Schaar, and R.~Leigh, {\it De {S}itter {H}olography and
  the {C}osmic {M}icrowave {B}ackground},  {\em JHEP} {\bf 0204} (2002) 047,
  [\href{http://xxx.lanl.gov/abs/hep-th/0202127}{{\tt hep-th/0202127}}].

\bibitem{larsen-mcnees}
F.~Larsen and R.~McNees, {\it Inflation and de {S}itter {H}olography},  {\em
  JHEP} {\bf 0307} (2003) 051,
  [\href{http://xxx.lanl.gov/abs/hep-th/0307026}{{\tt hep-th/0307026}}].

\bibitem{schaar}
J.~van~der Schaar, {\it Inflationary perturbations from deformed {CFT}},  {\em
  JHEP} {\bf 0401} (2004) 070,
  [\href{http://xxx.lanl.gov/abs/hep-th/0307271}{{\tt hep-th/0307271}}].

\bibitem{lidsey-liddle}
J.~Lidsey, A.~Liddle, E.~Kolb, E.~Copeland, T.~Barreiro, and M.~Abney, {\it
  Reconstructing the inflaton potential -- an overview},  {\em Rev. Mod. Phys}
  {\bf 69} (1997) 373, [\href{http://xxx.lanl.gov/abs/astro-ph/9508078}{{\tt
  astro-ph/9508078}}].

\bibitem{liddle-lyth}
A.~Liddle and D.~Lyth, {\em Cosmological Inflation and Large-Scale Structure}.
\newblock Cambridge University Press, Cambridge, 2000.

\bibitem{maldacena}
J.~Maldacena, {\it The large {N} limit of superconformal field theories and
  supergravity},  {\em Adv. Theor. Math. Phys.} {\bf 2} (1998) 231--252,
  [\href{http://xxx.lanl.gov/abs/hep-th/9711200}{{\tt hep-th/9711200}}].

\bibitem{witten-ads}
E.~Witten, {\it Anti de {S}itter {S}pace and {H}olography},  {\em Adv. Theor.
  Math. Phys.} {\bf 2} (1998) 253--291,
  [\href{http://xxx.lanl.gov/abs/hep-th/9802150}{{\tt hep-th/9802150}}].

\bibitem{aharony-gubser}
O.~Aharony, S.~Gubser, J.~Maldacena, H.~Ooguri, and Y.~Oz, {\it Large {N} field
  theories, string theory and gravity},  {\em Phys. Rept.} {\bf 323} (2000)
  183--386, [\href{http://xxx.lanl.gov/abs/hep-th/9905111}{{\tt
  hep-th/9905111}}].

\bibitem{johnson}
C.~Johnson, {\em {D}-{B}ranes}.
\newblock Cambridge University Press, Cambridge, 2003.

\bibitem{witten-ds}
E.~Witten, {\it Quantum {G}ravity in de {S}itter space},
  \href{http://xxx.lanl.gov/abs/hep-th/0106109}{{\tt hep-th/0106109}}. Prepared
  for International School of Subnuclear Physics: 39th Course: New Fields and
  Strings in Subnuclear Physics, Erice, Italy, 29 Aug -- 7 Sep 2001.

\bibitem{goheer-kleban}
N.~Goheer, M.~Kleban, and L.~Susskind, {\it The {T}rouble with de {S}itter
  {S}pace},  {\em JHEP} {\bf 0307} (2003) 056,
  [\href{http://xxx.lanl.gov/abs/hep-th/0212209}{{\tt hep-th/0212209}}].

\bibitem{dyson-kleban}
L.~Dyson, M.~Kleban, and L.~Susskind, {\it Disturbing {I}mplications of a
  {C}osmological {C}onstant},  {\em JHEP} {\bf 0210} (2002) 011,
  [\href{http://xxx.lanl.gov/abs/hep-th/0208013}{{\tt hep-th/0208013}}].

\bibitem{dyson-lindesay}
L.~Dyson, J.~Lindesay, and L.~Susskind, {\it Is {T}here {R}eally a de
  {S}itter/{CFT} {D}uality},  {\em JHEP} {\bf 0208} (2002) 045,
  [\href{http://xxx.lanl.gov/abs/hep-th/0202163}{{\tt hep-th/0202163}}].

\bibitem{klebanov-witten}
I.~Klebanov and E.~Witten, {\it Ad{S}/{CFT} {C}orrespondence and {S}ymmetry
  {B}reaking},  {\em Nucl. Phys. B} {\bf 556} (1999) 89--114,
  [\href{http://xxx.lanl.gov/abs/hep-th/9905104}{{\tt hep-th/9905104}}].

\bibitem{muck-viswanathan}
W.~M\"uck and K.~Viswanathan, {\it Regular and {I}rregular {B}oundary
  {C}onditions in the {A}d{S}/{CFT} {C}orrespondence},  {\em Phys. Rev. D} {\bf
  60} (1999) 081901, [\href{http://xxx.lanl.gov/abs/hep-th/9906155}{{\tt
  hep-th/9906155}}].

\bibitem{martelli-muck}
D.~Martelli and W.~M\"{u}ck, {\it Holographic {R}enormalization and {W}ard
  {I}dentities with the {H}amilton--{J}acobi {M}ethod},  {\em Nucl. Phys. B}
  {\bf 654} (2003) 248--276,
  [\href{http://xxx.lanl.gov/abs/hep-th/0205061}{{\tt hep-th/0205061}}].

\bibitem{bianchi-freedman-a}
M.~Bianchi, D.~Freedman, and K.~Skenderis, {\it How to go with an {RG} {F}low},
   {\em JHEP} {\bf 0108} (2001) 041,
  [\href{http://xxx.lanl.gov/abs/hep-th/0105276}{{\tt hep-th/0105276}}].

\bibitem{bianchi-freedman-b}
M.~Bianchi, D.~Freedman, and K.~Skenderis, {\it Holographic {R}enormalization},
   {\em Nucl. Phys. B.} {\bf 631} (2002) 159--194,
  [\href{http://xxx.lanl.gov/abs/hep-th/0112119}{{\tt hep-th/0112119}}].

\bibitem{muck-prisco}
W.~M\"{u}ck and M.~Prisco, {\it Glueball {S}cattering {A}mplitudes from
  {H}olography},  {\em JHEP} {\bf 0404} (2004) 037,
  [\href{http://xxx.lanl.gov/abs/hep-th/0402068}{{\tt hep-th/0402068}}].

\bibitem{bianchi-prisco-muck}
M.~Bianchi, M.~Prisco, and W.~M\"{u}ck, {\it New results on holographic
  three-point functions},  {\em JHEP} {\bf 0311} (2003) 052,
  [\href{http://xxx.lanl.gov/abs/hep-th/0310129}{{\tt hep-th/0310129}}].

\bibitem{haro-skenderis}
S.~de~Haro, K.~Skenderis, and S.~Solodukhin, {\it Holographic {R}econstruction
  of {S}pacetime and {R}enormalization in the {A}d{S}/{CFT} {C}orrespondence},
  {\em Commun. Math. Phys.} {\bf 217} (2001) 595--622,
  [\href{http://xxx.lanl.gov/abs/hep-th/0002230}{{\tt hep-th/0002230}}].

\bibitem{kalkkinen-martelli}
J.~Kalkkinen, D.~Martelli, and W.~M\"{u}ck, {\it Holographic renormalization
  and anomalies},  {\em JHEP} {\bf 0104} (2001) 036,
  [\href{http://xxx.lanl.gov/abs/hep-th/0103111}{{\tt hep-th/0103111}}].

\bibitem{fukuma-matsuura}
M.~Fukuma, S.~Matsuura, and T.~Sakai, {\it Holographic {R}enormalization
  {G}roup},  {\em Prog. Theor. Phys.} {\bf 109} (2003) 489--562,
  [\href{http://xxx.lanl.gov/abs/hep-th/0212314}{{\tt hep-th/0212314}}].

\bibitem{balasubramanian}
V.~Balasubramanian, P.~Kraus, and A.~Lawrence, {\it {B}ulk versus boundary
  dynamics in {A}nti-de {S}itter space-time},  {\em Phys. Rev. D} {\bf 59}
  (1999) 046003, [\href{http://xxx.lanl.gov/abs/hep-th/9805171}{{\tt
  hep-th/9805171}}].

\bibitem{skenderis}
K.~Skenderis, {\it Lecture {N}otes on {H}olographic {R}enormalization},  {\em
  Class. Quant. Grav.} {\bf 19} (2002) 5849--5876,
  [\href{http://xxx.lanl.gov/abs/hep-th/0209067}{{\tt hep-th/0209067}}].

\bibitem{nojiri-odintsov-holo}
S.~Nojiri and S.~Odintsov, {\it Conformal anomaly from d{S}/{CFT}
  correspondence},  {\em Phys. Lett. B} {\bf 519} (2001) 145--148,
  [\href{http://xxx.lanl.gov/abs/hep-th/0106191}{{\tt hep-th/0106191}}].

\bibitem{nojiri-odintsov-dscft}
S.~Nojiri and S.~D. Odintsov, {\it Asymptotically ds {S}itter dilatonic
  space-time, holographic {RG} flow and conformal anomaly from (dilatonic)
  d{S}/{CFT} correspondence},  {\em Phys. Lett. B} {\bf 531} (2002) 143--151,
  [\href{http://xxx.lanl.gov/abs/Text}{{\tt Text}}].

\bibitem{nojiri-odintsov-a}
S.~Nojiri and S.~Odintsov, {\it Quantum cosmology, inflationary brane-world
  creation and d{S}/{C}{F}{T} correspondence},  {\em JHEP} {\bf 0112} (2001)
  033, [\href{http://xxx.lanl.gov/abs/hep-th/0107134}{{\tt hep-th/0107134}}].

\bibitem{gubser-klebanov-polyakov}
S.~Gubser, I.~Klebanov, and A.~Polyakov, {\it {G}auge {T}heory {C}orrelators
  from {N}on-{C}ritical {S}tring {T}heory},  {\em Phys. Lett. B} {\bf 428}
  (1998) 105--114, [\href{http://xxx.lanl.gov/abs/hep-th/9802109}{{\tt
  hep-th/9802109}}].

\bibitem{balasubramanian-boer-a}
V.~Balasubramanian, J.~de~Boer, and D.~Minic, {\it Mass, {E}ntropy and
  {H}olography in {A}symptotically de {S}itter {S}paces},  {\em Phys. Rev. D}
  {\bf 65} (2002) 123508, [\href{http://xxx.lanl.gov/abs/hep-th/0110108}{{\tt
  hep-th/0110108}}].

\bibitem{balasubramanian-boer-b}
V.~Balasubramanian, J.~de~Boer, and D.~Minic, {\it Exploring de sitter space
  and holography},  {\em Class. Quant. Grav.} {\bf 19} (2002) 5655--5700,
  [\href{http://xxx.lanl.gov/abs/hep-th/0207245}{{\tt hep-th/0207245}}].

\bibitem{minces-rivelles}
P.~Minces and V.~Rivelles, {\it {E}nergy and the {A}d{S}/{C}{F}{T}
  {C}orrespondence},  {\em JHEP} {\bf 0112} (2001) 010,
  [\href{http://xxx.lanl.gov/abs/hep-th/0110189}{{\tt hep-th/0110189}}].

\bibitem{breitenlohner-freedman-a}
P.~Breitenlohner and D.~Freedman, {\it Stability in gauged extended
  supergravity},  {\em Ann. Phys.} {\bf 144} (1982) 249.

\bibitem{breitenlohner-freedman-b}
P.~Breitenlohner and D.~Freedman, {\it Positive energy in anti-de {S}itter
  backgrounds and gauged extended supergravity},  {\em Phys. Lett. B} {\bf 115}
  (1982) 197.

\bibitem{mezincescu-townsend}
L.~Mezincescu and P.~Townsend, {\it Stability at a local maximum in higher
  dimensional anti-de {S}itter space and application to supergravity},  {\em
  Ann. Phys.} {\bf 160} (1985) 406.

\bibitem{birrell-davies}
N.~Birrell and P.~Davies, {\em Quantum fields in curved space}.
\newblock Cambridge University Press, Cambridge, 1982.

\bibitem{morse-feshbach}
P.~Morse and H.~Feshbach, {\em Methods of Theoretical Physics}.
\newblock McGraw--Hill, 1953.

\bibitem{bianchi-marchetti}
M.~Bianchi and A.~Marchetti, {\it Holographic three-point functions: one step
  beyond the tradition},  {\em Nucl. Phys. B} {\bf 686} (2004) 261--284,
  [\href{http://xxx.lanl.gov/abs/hep-th/0302019}{{\tt hep-th/0302019}}].

\bibitem{peskin-schroeder}
M.~Peskin and D.~Schroeder, {\em An Introduction to Quantum Field Theory}.
\newblock Perseus Books, Reading, Massachusetts, 1995.

\bibitem{kogo-komatsu}
N.~Kogo and E.~Komatsu, {\it Angular {T}rispectrum of {CMB} {T}emperature
  {A}nisotropy from {P}rimordial {N}on-{G}aussianity with the {F}ull
  {R}adiation {T}ransfer {F}unction},
  \href{http://xxx.lanl.gov/abs/astro-ph/0602099}{{\tt astro-ph/0602099}}.

\end{thebibliography}
\end{document}